\documentclass[10pt,twocolumn]{article}

\usepackage[a4paper,margin=0.75in]{geometry}

\usepackage{tikz}
\usepackage{amsmath}

\usepackage{filecontents}

\usepackage{multirow}
\usepackage{tcolorbox}
\usepackage[utf8]{inputenc}
\usepackage{listings}
\usepackage{hyperref}
\usepackage{xcolor}
\usepackage{cleveref}
\usepackage{xcolor}
\usepackage{makecell}
\usepackage{subcaption}
\usepackage{subcaption}
\usepackage{booktabs}
\usepackage{xspace}
\usepackage[normalem]{ulem}
\usepackage{enumitem}
\useunder{\uline}{\ul}{}
\usepackage{url}

\newcommand{\tool}{CyberSleuth\xspace}
\newcommand{\baseline}{Single Agent\xspace}
\newcommand{\tshark}{Tshark Expert Agent\xspace}
\newcommand{\constrain}{Flow Reporter Agent\xspace}

\newcommand{\cfabench}{CFA-bench\xspace}

\definecolor{customgray}{gray}{0.7}
\definecolor{backcolour}{rgb}{0.95,0.95,0.92}
\definecolor{cmdcolour}{rgb}{0.9,1.0,0.9}

\lstnewenvironment{prompt}{%
    \thicklines
    \lstset{
        backgroundcolor=\color{backcolour},
        rulecolor=\color{backcolour!40!black},
        frameround=tttt,
        frame=single,
        basicstyle=\ttfamily\scriptsize,
        basewidth=0.50em,
        keywordstyle=\bfseries,
        xleftmargin=.02\linewidth,
        xrightmargin=.02\linewidth,
        escapeinside={<@}{@>},
        numbers=none
    }}{}

\lstnewenvironment{cmds}{%
    \thicklines
    \lstset{
        backgroundcolor=\color{cmdcolour},
        rulecolor=\color{cmdcolour!40!black},
        frameround=tttt,
        frame=single,
        basicstyle=\ttfamily\scriptsize,
        basewidth=0.50em,
        keywordstyle=\bfseries,
        xleftmargin=.02\linewidth,
        xrightmargin=.02\linewidth,
        escapeinside={<@}{@>},
        numbers=none
    }}{}

\begin{document}

\date{}
\title{
\tool: Autonomous Blue-Team LLM Agent for Web Attack Forensics
}

\author{
{\rm Stefano Fumero}\\
Politecnico di Torino
\and
{\rm Kai Huang}\\
Politecnico di Torino
\and
{\rm Matteo Boffa}\thanks{Corresponding author: \texttt{matteo.boffa@polito.it}}\\
Politecnico di Torino
\and
{\rm Danilo Giordano}\\
Politecnico di Torino
\and
{\rm Marco Mellia}\\
Politecnico di Torino
\and
{\rm Dario Rossi}\\
Huawei Technologies France
} 

\maketitle

\begin{abstract}
Post-mortem analysis of compromised systems is a key aspect of cyber forensics, today a mostly manual, slow, and error-prone task. Agentic AI, i.e., LLM-powered agents, is a promising avenue for automation. However, applying such agents to cybersecurity remains largely unexplored and difficult, as this domain demands long-term reasoning, contextual memory, and consistent evidence correlation -- capabilities that current LLM agents struggle to master.

In this paper, we present the first systematic study of LLM agents to automate post-mortem investigation. As a first scenario, we consider realistic attacks in which remote attackers try to abuse online services using well-known CVEs (30 controlled cases). The agent receives as input the network traces of the attack and extracts forensic evidence. 
We compare three AI agent architectures, six LLM backends, and assess their ability to i) identify compromised services, ii) map exploits to exact CVEs, and iii) prepare thorough reports.
Our best-performing system, \tool, achieves 80\% accuracy on 2025 incidents, producing complete, coherent, and practically useful reports (judged by a panel of 25 experts).
We next illustrate how readily \tool adapts to face the analysis of infected machine traffic, showing that the effective AI agent design can transfer across forensic tasks.
Our findings show that (i) multi-agent specialisation is key to sustained reasoning; (ii) simple orchestration outperforms nested hierarchical architectures; and (iii) the \tool design generalises across different forensic tasks.
\end{abstract}

\section{Introduction}
Cyber forensics is a cornerstone of modern cybersecurity operations. In particular, as cyber-threats continue to evolve, \textit{post-mortem analysis} of security incidents -- i.e., the systematic reconstruction of attacker's goals, actions, and impacted assets -- becomes essential to derive a \textit{lessons learned} and foster the improvement of defensive systems~\cite{nist80061r3}. 
Post-mortem analysis requires the security experts to i) extract relevant events from digital evidence, ii) correlate these events with existing knowledge bases, and iii) produce an analytical interpretation outlining the main components and interrelations of the incident. Today, the process remains highly manual, time-consuming, and prone to error~\cite{bridges2023testing}. With attacks becoming more sophisticated and automated~\cite{potter2025frontieraisimpactcybersecurity}, enhancing automation in forensic analysis is increasingly critical to keep defenders from falling behind.

Recent advances in LLMs have opened promising directions for automating reasoning-heavy tasks~\cite{AgentSurveyKDD25}, i.e., Agentic AI. In fact, LLM agents -- models equipped with structured reasoning loops, tool use, and memory -- have shown remarkable potential across domains, from software engineering to scientific discovery~\cite{chen2023program, schick2023toolformer, yao2023react}. In cybersecurity, most of the previous literature has focused on red-team applications, where agents are instructed to attack systems, find vulnerabilities, or generate exploits~\cite{deng2024pentestgpt, pentest_agent, wang2025cybergymevaluatingaiagents, zhu2025cvebench}. In contrast, defensive agents remain scarce~\cite{bui2024systematic, netmoonai, cfabench} and largely unexplored for cyber forensics~\cite{blue_red_team, potter2025frontieraisimpactcybersecurity}. Yet, designing LLM agents for forensic investigation is far from straightforward: forensic analysis demands sustained focus, consistent reasoning, and the ability to connect unstructured evidence across events. In reality, LLM agents struggle with memory management, context switching, and coordination among sub-agents. Therefore, it remains unclear whether they can effectively leverage forensic tools~\cite{AgentSurveyKDD25}. 

\begin{figure*}[thb]
    \centering
    \includegraphics[width=0.8\linewidth]{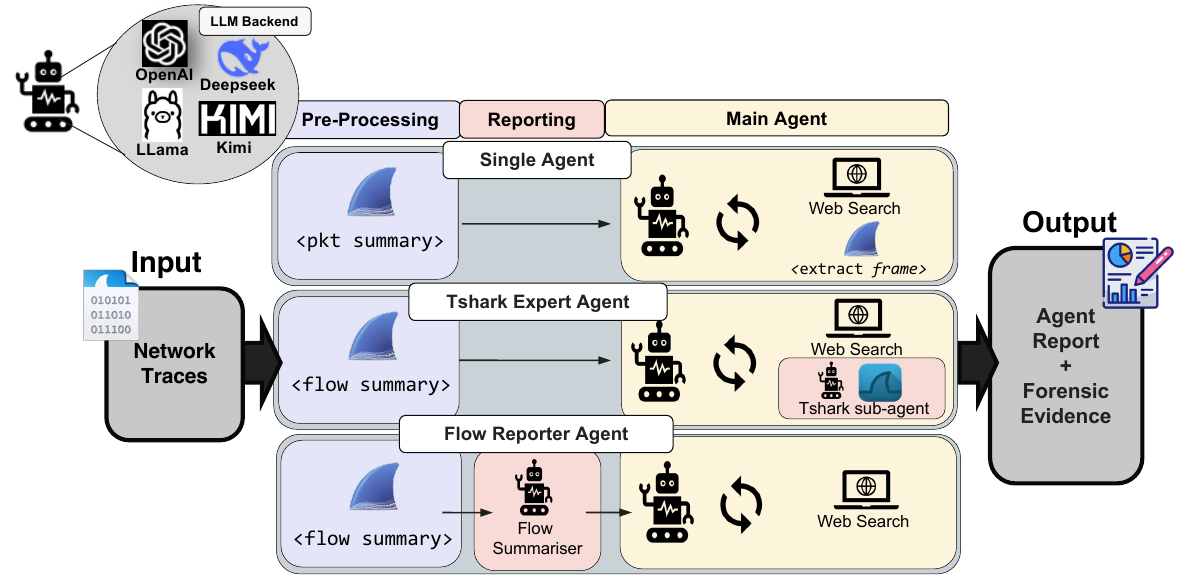}
    \caption{\textbf{Overview of agent architectures.} Each architecture receives the traffic trace and processes it through a pipeline of pre-processing, reporting, and reasoning. The agents connect to the desired LLM backend via API, interact with tools and sub-agents (\texttt{tshark}, Flow Summariser, and Web Search) to extract evidence and produce the forensic report.}
    \label{fig:overview_design_choices}
\end{figure*}

In this paper, we systematically study whether and how we could design an LLM agent to follow a forensic investigation pipeline.
We consider a web service, deployed within an enterprise network, that exposes possible vulnerable (or secure) services. The attacker -- who already knows the available endpoints -- runs a given exploit -- described by a Common Vulnerabilities and Exposures (CVE). Attacks against vulnerable services may succeed, while those against secure ones would fail. We assume a monitoring system that allows to record the network traces (in PCAP format), e.g., via a Web Application Firewall that terminates the TLS connection and allows dumping unencrypted packet traces.
We then instruct the LLM agent to run a post-mortem forensic investigation. The agent must autonomously analyse the traffic trace and (i) identify the targeted service, (ii) relate forensic evidence with the precise CVE the attacker exploited, and (iii) prepare a final report to summarise its findings, including whether the attack succeeded.
To evaluate different agent architectures and LLM, we leverage a benchmark of 30 controlled scenarios with increasing complexity, covering both legacy and recently disclosed vulnerabilities: we run the web service, conduct the attack, collect the trace. Being under our control, we know the ground truth, a fundamental need to support automated, fair and reproducible comparisons across various agent architectures and LLM backends~\cite{cfabench}.

We consider three agent architectures and six state-of-the-art LLM backends -- as shown in Figure~\ref{fig:overview_design_choices}. We build on a MemGPT-style~\cite{memgpt} multi-agent architecture, where specialised sub-agents collaborate toward a shared forensic goal.
We compare different solutions on the same playground: 20 out of 30 benchmark traces (with incidents up to 2024), digging into pitfalls and failure modes to identify the overall best-performing system, \tool. We next test it with previously unseen benchmark traces that include only 2025 vulnerabilities. \tool can correctly resolve 80\% of them, while producing reports that are, according to 25 experts, complete, useful and coherent. 
Beyond overall performance gains, we provide interpretable analyses that connect specific architectural choices to these improvements: effective prompt engineering, efficient memory management, and a multi-agent design all contribute to \tool’s success.

To evaluate \tool adaptability to other forensic tasks, we conduct a portability study and use the agent to analyse traffic generated by infected devices. We just task the agent with the new scenario (i.e., change the prompt) and keep the rest of the architecture unchanged. We test it with a public dataset of 10 traces referring to 10 different malware~\cite{malware_traffic_analysis}. 
\tool successfully extracts key information from malware traffic (e.g., victim details, Indicators of Compromise) and produces coherent and informative summaries.

The key takeaways from the design of \tool are:
\begin{itemize}
\item \textbf{Multi-agent design is essential}: a divide-and-conquer approach with specialised sub-agents outperforms a single, all-in-one agent, which tends to lose focus during complex analyses.
\item \textbf{Simple orchestration outperforms nested designs}: a sequential pipeline of agents proves more effective than deeply nested communication patterns, which often suffer from coordination failures.
\item \textbf{Best practices transfer across related tasks}: \tool performs well in attacks against web service and malware traffic analysis, showing that its design principles can be reused with minimal adaptation.
\end{itemize}

We release \tool and its benchmarks as an open platform to support rigorous evaluation.\footnote{\url{https://github.com/stefanofumero/LLM_Agent_Cybersecurity_Forensic/tree/main}}

\section{Related Works}
With the rapid progress of LLMs, researchers have increasingly investigated their potential to support security experts. Prior work mainly explored two directions: 
(i) \textit{blue-team agents}, using LLMs to defend systems by monitoring, reacting to, and reporting threats;
(ii) \textit{red-team agents}, simulating attacks to probe defences (e.g., penetration testing).
Our work contributes to the first line of research, while drawing methodological inspiration from the more mature literature on red-team agents~\cite{blue_red_team}.


\vspace{0.0cm}{$\bullet$ \textbf{Blue-team LLMs:}} A first line of research here consists on measuring whether the knowledge encoded in LLMs suffices for cyber-security tasks.
For instance, \textit{CSEBenchmark}~\cite{llms_security_experts} systematically evaluates 12 LLMs across 345 knowledge points expected from security experts. GPT- and DeepSeek-based models achieve the highest scores; yet all models perform poorly on \textit{procedural operations} (e.g., using advanced tools such as \texttt{tshark}). We observe the same limitation and propose mechanisms to mitigate it.  

Other studies focus on the forensic use of LLMs to aid security experts. One line of research uses LLMs to parse verbose CTI reports into structured outputs, e.g., knowledge graph extraction~\cite{ctinexus}, to identify Tactics, Techniques and Procedures (TTPs)~\cite{failure_location_llm}, or to generate incident reports -- either autonomously or in collaboration with experts~\cite{llms_security_incident}. Our setting is more demanding: our agent also needs to enrich and ground evidences (e.g., through web-search), rather than simply summarizing them.
Closer to our setting, a few works explore LLM-agents for network forensics~\cite{bui2024systematic,netmoonai,cfabench}. The agent in~\cite{bui2024systematic} is provided with pre-parsed payloads and metadata, whereas \tool autonomously extracts evidence from raw data. Similarly, NetMoonAI~\cite{netmoonai} generates qualitative outputs through a visual interface rather than structured forensic conclusions. CFABench~\cite{cfabench} is the most similar work to \tool; it introduces a benchmark for forensic agents (that we also compare with) and evaluates three ReAct-style designs. We extend their benchmark and fairly compare against their best-performing agent, improving CVE-retrieval accuracy from 14\% to 68\%. Overall, existing works are preliminary and lack the systematic design-space exploration that we perform.  
More generally, several papers investigate incident reporting and Root Cause Analysis (RCA) for server and cloud failures~\cite{comet,rca-agent_1,rca-agent_2,failure_location_llm,confucius}. These approaches share our idea of using agents to autonomously parse logs but address benign system failures, whereas our focus is on adversarial scenarios requiring cybersecurity expertise.

Other researchers explored the use of LLM-agents to actively solve or patch security incidents~\cite{nong2025appatch,log_in_patches_out,patchagent}. In \textit{PATCHAGENT}, researchers develop an agent which mimics how human experts triage bugs~\cite{patchagent}. \tool follows a similar intuition, as we design it to mimic the steps a human analyst would take. \textit{Nissist}~\cite{nissist} orchestrates multiple LLMs in a semi-autonomous loop (planner, retriever, etc.) to mitigate incidents. Unlike our work, the LLMs do not use tools and only process textual incident descriptions. Our agents must investigate directly from raw traces.  

\vspace{0.0cm}{$\bullet$ \textbf{Red-team LLMs:} Red-teaming agents gained traction early, led by \textit{PentestGPT}~\cite{deng2024pentestgpt}. We borrow some methodological insights from this line of work. For instance, \textit{AutoPenBench}~\cite{gioacchini2024autopenbench} introduces step-wise metrics to track agent progress during penetration tests. Following the same principle, our evaluation is \textit{sequential}: the agent cannot associate evidence with a vulnerability unless it has first identified the targeted service, nor can it assess attack success without first detecting the vulnerability.

Architectural innovations from red-team research also inform our design. \textit{Incalmo}~\cite{incalmo} separates high-level reasoning from low-level execution, improving performance in multi-host CTFs. Similarly, \textit{VulnBot}~\cite{kong2025vulnbot} orchestrates a team of specialised agents (planner, executor, analyser) to mitigate the “loss of context” faced by single-agent models. Our advanced multi-agent variants follow a similar rationale.

\vspace{0.0cm}{$\bullet$ \textbf{Benchmarking agentic solutions:} To conclude, we report recent papers highlighting ``best practises" when evaluating agents. \textit{From Promise to Peril}~\cite{blue_red_team} highlights the need for explainability, human oversight, and resilience. \textit{OCCULT}~\cite{occult} cautions against binary yes/no metrics that inflate performance. Finally,~\cite{best_practises_agents} proposes two principles for agentic benchmarks: \textit{outcome validity} (scores must reflect genuine success) and \textit{task validity} (tasks must only be solvable if the capability is truly present). Our benchmark explicitly adheres to these guidelines by avoiding binary metrics, excluding unsolvable tasks, and ensuring human-interpretable reasoning traces.

\section{Background on LLM Agents}\label{sec:Agents}
Here, we introduce some fundamentals on LLM agents. For more details, we refer the reader to~\cite{AgentSurveyKDD25}.

An \emph{agent} is an autonomous system that interacts with its environment to achieve specific goals. It collects inputs, processes information, and takes actions that influence the environment according to its objectives~\cite{sumers2023cognitive}. GenAI agents leverage the backend LLM model reasoning and planning abilities to guide decision-making.

The agent core components are:  
\begin{itemize}
 \item \textbf{Reasoning \& Decision-Making:} The LLM performs iterative reasoning steps, analyses observations, formulates intermediate thoughts, and decides the next action.
 \item \textbf{Actions:} Allow the agents to interact with the environment through predefined \emph{tools}, i.e., executing applications, querying APIs, or updating internal memory.
 \item \textbf{Memory:} Agents use short-term memory for immediate interactions, and long-term memory to retain past knowledge.
\end{itemize}

The agent is programmed via an initial \textit{prompt} that specifies the task, available tools, and input data (which accumulate over interactions). The agent reasons about which actions to take, executes them, and observes the responses of the environment. Through iterative cycles of perception, reasoning, and action, the LLM agent can autonomously solve complex tasks that could not be scripted manually.

Importantly, the prompt size is limited by the LLM’s Context Window\footnote{Number of tokens (text units) that the model can process at once.}, which bounds how much text (thoughts, actions, observations, and memory) the model can handle at once. In most current agent systems, memory is therefore short-term by design: the agent’s working memory is simply what fits inside the prompt. This approach makes agents easy to implement but inherently forgetful -- earlier context can get lost.
Long-term memory, on the contrary, is not an intrinsic feature of LLM agents. It requires explicit external components -- such as vector databases or document stores -- to persist and retrieve information across interactions~\cite{AgentSurveyKDD25}. Implementing such mechanisms is non-trivial and remains an open challenge for reliable long-term reasoning.

LLM-powered systems can be extended to \emph{multi-agent} architectures, where multiple \textit{sub-agents} specialise in different aspects of a task and collaborate, coordinated by a \textit{main agent}. Each agent may specialise in a specific role, e.g, one dedicated to evidence analysis, another to web search, and a third to report generation. Multi-agent setups enable parallel reasoning, reduce cognitive load on individual agents, and can improve robustness by cross-validating observations and decisions. Their orchestration becomes critical.

\section{Threat Models: Definitions}\label{sec:ThreatModel}


We introduce two \textit{threat models} and scenarios: i) attacks against web services and ii) traffic analysis of malware-infected devices. 
We provide a summary in Table~\ref{tab:schema_threat_model}.

\begin{table*}[t]
\centering
\caption{\textbf{Threat models for the security scenarios we consider}. For each threat model, we identify the \textit{threat actor} (the entity performing the attack), \textit{attack vectors} (the methods or paths used to carry out the attack), \textit{threat goals} (the attacker’s intended outcomes), and \textit{assumptions} (the environmental conditions and defender capabilities assumed in the scenario).}
\label{tab:schema_threat_model}
\resizebox{\textwidth}{!}{%
\begin{tabular}{p{3cm}p{3cm}p{3.5cm}p{3cm}p{3.5cm}}
\toprule
\textbf{Scenario} & \textbf{Threat Actor} & \textbf{Attack Vectors} & \textbf{Threat Goals} & \textbf{Assumptions} \\
\midrule
Attacks Against Web Services & Skilled adversaries perform targeted exploitation of exposed web applications & Crafted requests exploiting known vulnerabilities (e.g., CVEs) & Privilege Escalation, System Compromise, Data Exfiltration, DoS & System monitors in-clear traffic via TLS termination and selective pcap logs; we directly observe exploitation. \\
\midrule
Traffic analysis of Malware-Infected Devices & Financially motivated cybercriminals leveraging commodity malware & Malicious executable distri-buted via phishing, malicious downloads, remote control channels, remote exploit & Long-term observation, Cryptomining, Botnet recruitment & Compromised device traffic passes through a monitoring node that logs encrypted and cleartext traffic. \\
\bottomrule
\end{tabular}}
\end{table*}


\vspace{0.1cm}$\bullet$ \textbf{Attacks against Web Services}: The adversary tries to compromise a web application by exploiting known vulnerabilities (e.g., CVEs) in exposed services. Attackers are aware of reachable endpoints and craft arbitrary application-layer requests to run the exploit. The attackers have multiple threat goals, e.g., remote code execution, privilege escalation, system compromise, data exfiltration, or launching denial-of-service attacks. The enterprise network deploys inspection components such as \textit{Web Application Firewall} or \textit{TLS termination points} that make it possible to log cleartext network traffic for later analysis (but cannot prevent the attack). We do not consider large-scale scanning and automated discovery phases; the focus is on targeted exploitation attempts. Attacks may succeed or fail depending on service vulnerabilities. All relevant network traces are available for post-mortem forensic reconstruction. For simplicity, we assume traffic not related to the attack is already filtered.

\vspace{0.1cm}$\bullet$ \textbf{Traffic of Malware-Infected Devices}: An adversary compromises a device and installs a \textit{commodity malware}. The initial infection typically occurs through phishing, malicious downloads from compromised websites, remote exploits, etc. We observe the initial exploitation phase (malware installation) and the subsequent phases when the malware is active: the communications with some Command and Control node and the Action on Objective, e.g., reporting long-term observations via keyloggers, controlling installed cryptomining tools, managing the node as part of botnets, etc. All network communications pass through a monitoring node, which records encrypted and clear-text traffic exchanged by the attacked device. This traffic provides observable evidence for post-infection forensic analysis.
This scenario includes the case where the security analyst intentionally runs malware on a controlled machine while it records the traffic it generates.

In both scenarios, our forensic agent receives the captured PCAP traces as the only input and performs automated post-mortem analysis, such as reconstructing the attack timeline and generating structured incident reports.

\section{Designing Agents for Forensic Analysis}\label{sec:tool}

In the following sections, we present our methodology for designing LLM agents for forensic analysis. Using the scenario of attacks against web applications, we develop three distinct architectures to systematically study how different design choices, such as input data, tool integration, and multi-agent organisation, affect forensic performance. Later, we face the analysis of malware-generated traffic to gauge the agent portability.

We start by describing the common components of all agents. We highlight \colorbox{backcolour}{prompts} and \colorbox{cmdcolour}{shell commands.}

The agent takes as input the packet traces recorded during the attack and must return (i) the targeted service, (ii) the CVE exploited by the attacker, (iii) an assessment of whether the attack succeeded, and (iv) a structured forensic report documenting the findings.
Referring to Fig.~\ref{fig:overview_design_choices}, agents share a common three-stage pipeline -- \textit{pre-processing}, \textit{main agent execution}, and \textit{reporting}. All agents adopt the same memory management mechanism, task prompt, and access to a Web Search tool. We implement all agents using LangChain~\cite{langchain_docs} and LangGraph~\cite{langgraph}, a graph-based framework for agent development. In practice, LangGraph models the execution flow as a dynamic graph in which the agent decides whether to continue reasoning, invoke a tool, or finalise the report. Graph nodes represent reasoning steps or tool invocations. After each step, the agent updates its memory and uses it to inform subsequent decisions.

\subsection{Task and Prompt}\label{sec:task_prompt}

Following good practice, our prompts always contain:

\begin{itemize}
    \item \textbf{A description of the agent's role}: describes the scenario and invite the agent to think step-by-step~\cite{wang2023selfconsistency}.
    \item \textbf{System instructions and available tools}: lists the execution rules (e.g., provide one reasoning step at a time) and the available tools (e.g., web-search).
    \item \textbf{Goals and guidelines}: specifies what the agent shall return as output and static suggestions we want our agent to remember.
\end{itemize}

In the following, we report the Goals and guidelines for the scenario of attacks against web applications (full prompt in the Appendix~\ref{sec:prompts}):

\begin{prompt}
GOALS: Analyse the provided PCAP (Packet Capture) file 
to accomplish the following tasks:
1. Identify the name of the service or application involved.
2. Gather evidence of malicious activities.
3. Relate the evidence with existing CVEs (Common 
Vulnerabilities and Exposures) and pick the most relevant.
4. Assess whether the service or application is vulnerable to 
the identified attack.
5. Evaluate whether the attack was successful.

GUIDELINES: Begin by thoroughly analysing the PCAP file to 
extract relevant information. After the exploratory analysis, 
ensure that the CVE identification is accurate by 
cross-referencing details from external sources with the 
evidence found in the PCAP files. Use the online search tool
only after the exploratory analysis has been completed to 
verify the findings and gather additional information.
\end{prompt}

We refined and improved prompt wording for conciseness and clarity, ensuring that the agent receives explicit, structured instructions while minimising ambiguity. We found the ``prompt engineering'' important, but not a fundamental part of the agent design.

\subsection{Long-term Memory Management}

As discussed in Section~\ref{sec:Agents}, memory management is a critical component of LLM-based agents. During execution, the agent must be aware of past actions, reasoning and their outcomes. A simple approach -- used in prior work~\cite{cfabench} -- is to rely on the \emph{ReAct} (Reasoning and Acting) framework~\cite{yao2023react}. In this setup, the agent records each observation, action, and result on a \textit{scratchpad}, which is continuously updated through the reasoning process. The scratchpad, appended to the original prompt, serves as a short-term memory that enables the agent to track its progress. However, as it is part of the prompt, it consumes part of the LLM’s context window. Consequently, it can quickly exhaust the available context, forcing older information to be discarded\footnote{Excessive summarisation also leads to degraded performance~\cite{evolving_contexts}.}.

To overcome this limitation, we take inspiration from the \emph{MemGPT} memory management framework~\cite{memgpt}, which augments short-term context with an external long-term memory. Building on this idea, we introduce a \textit{vector database} that stores past experiences outside the LLM’s prompt. The database acts as a key–value memory: each experience deemed relevant by the agent is summarised, embedded (i.e., mapped into a vector space), and stored -- using the embedding as the key and the summary as the value. During subsequent reasoning, the agent gets past experiences whose embeddings are semantically similar to its current context, recalling prior in-context knowledge when needed.
At each execution step, the agent operates with three complementary memory components:
\begin{itemize}
\item \textbf{System Instructions}: Static definitions of the agent’s role, available tools, and assigned tasks.
\item \textbf{FIFO Queue}: A token-limited buffer containing the most recent actions and observations. It represents the agent’s short-term memory, analogous to the traditional scratchpad.
\item \textbf{Working Context}: The dynamic long-term memory component containing the top-10 most relevant past events retrieved from the vector database.
\end{itemize}
To populate the working context, the agent computes embeddings of the latest FIFO entries and retrieves semantically similar records from the vector database. This mechanism enables the agent to maintain coherence, recall relevant past insights, and avoid redundancy during long forensic investigations.

\subsection{Tool: Focussed Web Search}

Considering tools that agents can use, we engineer a specialised web-search tool to enrich and ground their findings. This process mimics the work of security experts when correlating events with existing knowledge bases. In particular, in the framework of attacks against web applications, the web-search tool links the forensic evidence with known vulnerabilities (CVEs). The main agent formulates queries in natural language, which are then executed by the web search tool. When constructing the queries, we explicitly suggest the LLM to (i) not include the CVE code (\texttt{CVE-XXXX-YYYY}) of specific vulnerabilities (even larger models hallucinate these codes); (ii) include the identified web service and type of attack (to push the model to really interpret the forensic evidence).

We use the Google Custom Search API~\cite{googlecse} to implement the web-search tool. An auxiliary sub-agent retrieves the top 10 results. It then visits and summarises each page and collates these summaries into a single search report. The sub-agent returns its report to the main agent, which can then cross-check and consolidate the forensic evidence from past findings and the network traces. 
Compared with past researchers~\cite{cfabench}, that used standard web search tools, we found that summarising and collating results to produce concise reports is critical to maintain the main agent's focus: Short and structured summaries let the agent reason more reliably and efficiently. 

\section{Agent Architectures}

Figure~\ref{fig:overview_design_choices} summarises the three agent alternatives and highlights the main differences between them. All agents share the same prompting schema, memory management, and web search tool previously described. Their main differences lie in how they access and browse the input data collected during the incident. We provide the prompt for each agent and sub-agent in Appendix~\ref{sec:prompts}. For each alternative, we provide a brief description, the available tools, and the operational workflow (sequence of actions the agent can perform).

\subsection{\baseline (SA)}
\label{sec:AgentSA}

The first architecture adopts the same single-agent architecture introduced in CFABench~\cite{cfabench}. 
Given the \textit{list of available packets}, the agent interacts with the environment and external sources through the \emph{PCAP Reader} tool and the \emph{Web Search Tool}. 

\vspace{0.1cm} \textbf{Tool -- PCAP Reader}: It is a simple wrapper around \texttt{tshark} that lets the agent get the application payload information of a desired packet:
\begin{cmds}
tshark -r {pcap_file} -Y "frame.number=={frame_number}" 
-T fields -e data
\end{cmds}

\vspace{0.1cm} \textbf{Operational Workflow}: Shown in the top block of Figure~\ref{fig:overview_design_choices}, we populate the SA agent memory with the \textit{packet list summary} of the entire trace obtained by running \texttt{tshark}:
\begin{cmds}
tshark -r {pcap_file} -T fields -e frame.number -e frame.time 
-e frame.protocols -e _ws.col.Info
\end{cmds}
This summary lists all packets, including Frame ID, timestamp, highest-level protocol, and a concise information field.
The SA analyzes this summary to autonomously identify packets of interest and retrieves their payloads through the PCAP Reader tool. The \emph{Web Search Tool} allows the agent to access external knowledge. Intermediate reasoning results are stored in the FIFO queue to preserve short-term context across steps and progressively consolidated them into long-term memory. The process iterates until the forensic task completes or the maximum number of iterations is reached. 

\baseline emulates the reasoning process of an experienced security analyst: a single glance at the packet summary suffices to pinpoint where to look for key evidence (i.e., which packets to inspect with the PCAP Reader). The intuition is that, if this architecture performs effectively, current LLM-based agents already demonstrate surprisingly advanced forensic capabilities.

\subsection{\tshark (TEA)}\label{sec:AgentTSE}

In our second proposal, TEA, we take a step back and make more cautious assumptions.
Rather than considering the packet list, we offer the agent a concise view of the \textit{TCP and UDP flow list}. This flow-based perspective reduces the input size while preserving the essential structure of the communication -- particularly useful when attacks involve hundreds of thousands of packets, causing the packet compact summaries to exceed the LLM’s context window.
We then enable the agent to further analyze individual network flows. To this end, we introduce a \textit{dedicated tshark sub-agent} that offloads packet inspection from the main agent, allowing it to remain focused on reasoning tasks. This division of roles improves command precision, reduces syntax errors, and avoids irrelevant operations, resulting in a more reliable workflow.

The tshark sub-agent (i) receives high-level natural language instructions from the main agent, (ii) translates them into actual tshark commands, and (iii) returns concise summaries of the extracted evidence. The main agent can run to the Web Search Tool to gather complementary external information. 

\vspace{0.1cm} \textbf{Tool -- Tshark Sub-agent}:
We implement the \emph{Tshark sub-agent} with a modular structure, mirroring the design of the main agent. Upon receiving a natural language instruction, it uses two tools: the \emph{Tshark Manual Reader} and the \emph{Tshark Executor} tools

\begin{itemize}
    \item \textbf{Tshark Manual Reader} is a semantic search tool built from the official \texttt{tshark} documentation~\cite{tsharkManpage} and filter guide~\cite{wiresharkFilterManpage}. It allows the sub-agent to retrieve the most relevant manual sections given the main agent’s instruction, ensuring accurate and context-aware command generation.
    \item \textbf{Tshark Executor} is a command execution tool that runs the generated \emph{tshark} commands, analyses the outputs, automatically detects and corrects errors, refines filters, and truncates results to respect constraints.
\end{itemize}
The sub-agent iteratively refines and executes commands, producing a structured summary once the analysis is complete (or the maximum iteration threshold is reached).

\vspace{0.1cm} \textbf{Operational Workflow}:
As shown in the second block of Figure~\ref{fig:overview_design_choices}, TEA receives the PCAP trace and the \textit{TCP and UDP flow list} extracted using:

\begin{cmds}
tshark -r <pcap_file> -q -z conv,tcp -z conv,udp
\end{cmds}
whose output is included with TEA’s initial prompt before the reasoning process begins, during which TEA defines which flows necessitate deeper inspection. It delegates this task to the \emph{Tshark sub-agent} via a natural language instruction, for instance:

\begin{prompt}
Follow flow n. {XX} in the PCAP file, report relevant details.
\end{prompt}
The \emph{Tshark Agent} interprets the request, supported by the \emph{Manual Reader Tool}. It generates and executes the appropriate commands with the \emph{Executor Tool}, and finally returns a concise summary of the findings along with the executed command history for completeness.

The main agent is responsible for orchestrating the web search tool, the \emph{Tshark Sub-agent} tools, and performing its memory management. This creates a multi-agent architecture which enables complex interactions.

\subsection{\constrain (FRA)}\label{constrained-agent}

The complex orchestration of the TEA multi-agent architecture may limit its efficiency. For instance, the main agent may issue overly broad requests to the \textit{Tshark Sub-agent}, which in turn may produce low-quality responses that call for more analysis, causing the overall interaction to diverge. To address this, we introduce \constrain. FRA adopts a simple \textit{sequential multi-agent system} design: the agent first inspects the PCAP files to extract relevant evidence; then performs an in-depth analysis of the results to detect anomalies. 

To this end, we couple the main agent, in charge of interpreting the results and enriching them with online references, with a \emph{Flow Summariser sub-agent}. This sub-agent systematically inspects TCP/UDP flows in the input file and generates a structured report summarising the most important forensic indicators (unusual ports, suspicious payload, or suspicious timing patterns). This enables a thorough traffic inspection and provides a targeted context for the main agent. Eventually, the main agent analyses the report and performs the forensic investigation, relying only on the \emph{Web Search Tool}.

\vspace{0.1cm} \textbf{Tool -- Flows Summariser Sub-agent}: is a dedicated agent that is proficient in inspecting trace files. For each TCP/UDP flow in a PCAP file, it reconstructs the complete application-layer exchange using \emph{tshark} to extract the ASCII payload and create a text representation:
\begin{cmds}
tshark -r <pcap_file> -q -z follow,{tcp|udp},ascii,{flow_number}
\end{cmds}
Armed with all flow data, we task the \textit{Flows Summariser sub-agent} to generate a structured forensic report describing i) \textbf{service(s)} involved and, if possible, the version to support accurate web queries; ii) \textbf{relevant events}, including the IP addresses involved and, optionally, quotes the content of the traffic to support interpretation; iii) \textbf{suspect malicious activities}, coupled with the corresponding service; and iv) \textbf{evidence} of whether the attack succeeded.

To save tokens and time, we avoid calling the \textit{Flows Summariser sub-agent} for TLS-encrypted flows because they contain little information and can hinder the analysis. Also, given that the payload of a flow can quickly become very large and overflow the LLM context window, we introduce a constraint to control the amount of payload to process. Given an estimate of the total number of tokens that the entire payload would consume, we apply a token allocation strategy based on square root allocation. 

For each flow $i$, compute:

\begingroup
\setlength{\abovedisplayskip}{4pt}
\setlength{\belowdisplayskip}{4pt}
\[
\begin{aligned}
\text{allocation}_i &= \text{budget}
   \frac{\sqrt{\text{tokens}_i}}{\sum_j \sqrt{\text{tokens}_j}}, \\
\text{final\_allocation}_i &= \min(\text{tokens}_i, \text{allocation}_i)
\end{aligned}
\]
\endgroup

%

This heuristic prevents large flows from consuming a disproportionately large part of the context budget and, at the same time, ensures that smaller but potentially important flows are analysed.\footnote{We tested block chunking, overlapping-window chunking, and a linear allocation strategy. The root-square allocation proved to be the most robust.} Specifically, the \emph{Flows Summariser Agent} allocates half of the tokens to the beginning and half to the end of the flow. This strategy increases the likelihood of capturing important forensic indicators that are typically located in session boundaries, such as attack signatures or service banners.

\vspace{0.1cm} \textbf{Operational flow}:
As shown in the third block of Figure~\ref{fig:overview_design_choices}, FRA receives as input the same \textit{TCP and UDP flow list} provided to TEA. The \textit{Flow Summariser sub-agent} gets this list and inspects each flow individually to create its summary. The main agent receives the summary and proceeds with its forensic investigation, reasoning, using the \emph{Web Search Tool}, and managing its memory.

\section{Datasets and Metrics}\label{sec:benchmark}

In this section, we describe the datasets and metrics for our cybersecurity scenarios. 

\begin{table*}[tb]
\centering
\caption{\textbf{Incidents for scenario of attacks against web applications}. Upper incidents are sorted according to the ability of the agents to solve them -- \textbf{simpler first}. Bottom ones are the \textbf{new 2025 incidents} we collected.}
\label{tab:benchmarks}
\footnotesize
\begin{tabular}{llccccccc}

\toprule
\textbf{ID} & \textbf{Service}            & \textbf{Version}  & \textbf{CVE}            & \textbf{\# Containers} & \textbf{Success} & \textbf{Pkts} & \textbf{Volume} & \textbf{Log \# Lines}\\ 
\midrule
0 & Apache Solr        & 8.11.0   & CVE-2021-44228  & 1       & True   & 22 &    2.89KB     &  $156^{\dagger\star}$ \\
1 & Jenkins            & 2.441    & CVE-2024-23897  & 1       & True   & 798 &    2.95MB    &  $3^{\dagger}$\\
2 & SaltStack          & 3002     & CVE-2020-11651 & 1       & False  & 29 &    2.66KB      &  $1^{\star}$\\
3 & Grafana            & 8.2.6    & CVE-2021-43798  & 1       & True   & 276 &    721.73KB  &  $387^{\dagger\star}$\\
4 & Apache HTTP Server & 2.4.49   & CVE-2021-41773  & 1       & True   & 113 &    13.30KB   &  $11^{\star}$ \\
5 & SaltStack          & 2019.2.3 & CVE-2020-11651 & 1       & True   & 32 &    2.42KB      &  $18$\\
6 & Apache HTTP Server & 2.4.50   & CVE-2021-42103  & 1       & False   & 114 &    13.06KB  &  $11^{\star}$ \\
7 & Apache ActiveMQ    & 5.14.2   & CVE-2017-15709 & 1       & True   & 18 &    1.54KB      & $40^{\dagger}$ \\ 
8 & Apache ActiveMQ    & 5.17.3   & CVE-2017-15709 & 1       & False  & 12 &    1.08KB      & $44^{\dagger}$ \\ 
9 & Couchdb            & 3.2.1    & CVE-2022-24706  & 1       & True   & 35 &   2.61KB      & $1^{\dagger}$\\ 
10 & phpMyAdmin        & 4.4.15.6 & CVE-2016-5734   & 1       & True   & 280 &   65.46KB    & $4^{\dagger\star}$ \\
11 & phpMyAdmin        & 4.8.1    & CVE-2018-12613  & 1       & True   &  327 &    80.87KB  & $5^{\star}$     \\
12 & Apache APISIX     & 2.9      & CVE-2021-45232 & 2       & True   & 390 &    50.62KB    & $108^{\dagger\star}$\\
13 & Joomla            & 4.2.7    & CVE-2023-23752  & 1       & True   & 596 &    205.09KB  & $14$\\
14 & GitLab            & 13.10.0  & CVE-2021-22205 & 3       & True   & 6680 &    19.10MB   & $190^{\dagger}$\\
15 & GitLab            & 13.10.3  & CVE-2021-22205 & 3       & False  & 9978 &    25.50MB   & $170^{\dagger}$\\
16 & Apache Airflow    & 1.10.10  & CVE-2020-11981 & 7       & True   & 793 &    174.13KB   & $967^{\dagger\star}$\\ 
17 & Cacti             & 1.2.22   & CVE-2022-46169  & 1       & True   & 2028 &    257.47KB & $15^{\dagger}$\\
18 & Apache Airflow    & 1.10.10  & CVE-2020-11981 & 7       & False  & 131 &    34.55KB    & $514^{\dagger}$\\ 
19 & Apache APISIX     & 2.11     & CVE-2021-45232 & 3       & False  & 500 &    74.03KB    & $99^{\dagger}$\\
\midrule
\midrule
20 & Vite        & 6.2.2 & CVE-2025-30208 & 1  & True  & 24 & 8.20KB    & $13^{\dagger}$\\
21 & Vite        & 6.2.3 & CVE-2025-30208 & 1 & False & 24 & 3.48KB     & $14^{\dagger}$\\
22 & Next.js     & 15.2.2 & CVE-2025-29927 & 1  & True  & 26 & 11.54KB  & $9^{\dagger}$\\
23 & Next.js     & 15.2.3 & CVE-2025-29927 & 1 & False & 24 & 2.46KB    & $10^{\dagger}$\\
24 & Langflow    & 1.2.0 & CVE-2025-3248  & 1  & True  & 20 & 2.18KB    & $18^{\dagger}$\\
25 & Langflow    & 1.2.0 & CVE-2025-3248  & 1  & True  & 135 & 121.01KB & $18^{\dagger}$\\
26 & Langflow    & 1.5.0 & CVE-2025-3248  & 1 & False & 44 & 6.23KB     & $405^{\dagger}$\\
27 & Tomcat      & 9.0.97& CVE-2025-24813 & 1  & False & 54 & 29.59KB   & $40^{\dagger}$\\
28 & Erlang/OTP  & 27.3.2& CVE-2025-32433 & 1  & True  & 20 & 2.97KB    & $2$\\
29 & Erlang/OTP  & 27.3.3& CVE-2025-32433 & 1 & False & 19 & 2.88KB     & $2$\\
\bottomrule
\end{tabular}

\footnotesize
[\dag] Logs contain information about service. [$\star$] Logs contain information about attack patterns or outcome.
\end{table*}

\subsection{Attacks Against Web Applications}

We use this threat model to compare different agent architectures and select the best-performing agent as \tool. We report details in Tab.~\ref{tab:benchmarks}. 

\vspace{0.1cm} \textbf{Dataset -- Agent Design}: We build on \cfabench~\cite{cfabench}. The original benchmark consists of 20 web-based incidents targeting containerised services collected in a controlled testbed. Each incident involves an external attacker trying to exploit a service specific CVE. 15 of these attacks are conducted by malicious HTTP requests (REST-API, buffer overflow, etc.). Others involve non-web applications like databases, middleware, and systems involving multiple components. 14 attacks succeeded (vulnerable system), while the 6 failed (either patched systems, or vulnerable component disabled). The benchmark provides network traces for each incident. The traces are not encrypted and include only traffic between the service and the attacker. The trace size of different incidents varies greatly, ranging from a dozen to thousands of packets and megabytes of data. We use these 20 incidents to systematically pick design choices when building our agent. 
For completeness, we collect and examine the application logs associated with each incident -- which we do not feed to our agent -- to assess their diagnostic value. These logs contain service-level information in 22 cases and explicit traces of the attack vector in only 9 of them, indicating that, for this scenario, application logs alone are insufficient for post-mortem reconstruction. Conversely, the corresponding PCAP traces consistently include clear forensic evidence of the attack in all analysed cases.

\vspace{0.1cm} \textbf{Dataset -- Agent Testing}: We expand the original dataset by adding 10 new incidents for independent evaluation. To test agents on scenarios involving information published after LLMs’ training, we include 5 services (4 HTTP-based and 1 SSH-based) affected by newly disclosed vulnerabilities in 2025. For each service, we deploy both vulnerable and patched versions and carry out attacks against both.

\vspace{0.1cm} \textbf{Dataset -- Benign Browsing}: For completeness, we collect 10 traces of routine web-browsing activity (e.g., logins to popular services, web searches, and general browsing). These sessions contain no attacks; in two login attempts, we intentionally entered a few incorrect passwords to simulate a user forgetting their credentials.\footnote{We collect traces enabling the \texttt{SSLKEYLOGFILE} to collect the TLS keys and passing those to the agents to decrypt the payload.} We use this benign data to evaluate how \tool handles normal traffic and whether it produces false positives.

\vspace{0.1cm} \textbf{Quantitative Metrics}: We consider the \cfabench metrics. The agent output must include 3 \textit{checkpoints} to assess its correctness. The checkpoints -- true/false tests -- measure the ability of the agent to identify i) the service under attack (\textbf{Service Identification}); ii) the specific vulnerability/CVE code (\textbf{CVE detection}); and iii) whether the attack was successful (\textbf{Attack Success Evaluation}). As the dataset is imbalanced (14 successful vs. 6 failed attacks in the Agent Design dataset), we also report the F1 score and the Matthews Correlation Coefficient (MCC)~\cite{MCC1975}\footnote{Ranges between $[-1, 1]$; offers a balanced measure of performance under class imbalance. $1$ indicates perfect prediction; $0$ random guessing; $-1$ total disagreement between predictions and ground truth}. We further report \textbf{internal agents metrics} to record the number of reasoning steps before termination, the number of input and output tokens processed by the LLM, and the associated monetary cost for each run (obtained from the LLM platform directly). Note that tokens directly affect the cost, with input tokens being less expensive than output tokens for cloud-based LLMs.

\vspace{0.1cm} \textbf{Human Evaluation of Forensic Report}: Since we cannot provide quantitative metrics on the full forensic report, we recruited human experts to perform a qualitative assessment of the agents’ performance. Experts were asked to rate, on a scale from 0 to 5 (0 = Totally Unsatisfied, 5 = Totally Satisfied), whether the report: (i) includes relevant and accurate information (\textbf{Completeness}), (ii) is helpful to a human analyst (\textbf{Usefulness}), and (iii) presents clear and logical reasoning (\textbf{Logical Coherence}). We collected responses from 25 volunteers, evaluating four incidents from the 2025 test set for which the agent completed the investigation successfully. Each participant received the agent’s reasoning steps, the final report, and the corresponding PCAP file. Volunteers also self-reported their level of forensic expertise before completing an online questionnaire.\footnote{The questionnaire and reports are available \href{https://forms.gle/QsaJ1dVUSK7suu3K6}{here}}

\subsection{Analysis of Malware Traffic}

We use this threat model for a \textit{portability} study to evaluate \tool in a different but related forensic scenario. 

\vspace{0.1cm} \textbf{Dataset -- Traffic Exercises}: We build on the publicly available malware traffic traces~\cite{malware_traffic_analysis}. This dataset is maintained by a principal threat researcher at Palo Alto Networks Unit 42, who curates and publishes real-world PCAP traces from observed malware infections. The traces come from controlled environments. The researcher intentionally infects Windows hosts with various malware samples and records the traffic observed during the infection and subsequent activity. Each trace comes with contextual information describing the infection scenario, victim system, and indicators of compromise (IOCs) such as malicious domains and IP addresses. We selected 10 recent traces covering a diverse set of malware families and infection vectors. We report details in Table~\ref{tab:malware_dataset}. Traces, containing a larger number of packets with respect to the previous scenario, contain both benign background traffic from the victim host and malicious communications -- such as command-and-control exchanges, payload downloads, or data exfiltration. 

\begin{table}[tb]
\centering
\footnotesize
\setlength{\tabcolsep}{3.5pt}
\renewcommand{\arraystretch}{1.05}
\caption{\textbf{Incidents for scenario of malware traffic.} Captures record the malware infection and subsequent exploitation, along with background traffic. Some incidents involve infection chains with multiple related malware components.}
\label{tab:malware_dataset}
\begin{tabular}{@{}lccc@{}}
\toprule
\textbf{Event} & \textbf{Malware} & \textbf{Pkts} & \textbf{Volume} \\
\midrule

\makecell[l]{\href{https://www.malware-traffic-analysis.net/2025/01/22/index.html}{Fake Download}\\(2025)} 
& Loader, Infostealer & 39k & 26.80MB \\

\makecell[l]{\href{https://www.malware-traffic-analysis.net/2024/11/26/index.html}{Nemotodes}\\(2024)} 
& NetSupport RAT & 27k & 21.30MB \\

\makecell[l]{\href{https://www.malware-traffic-analysis.net/2024/09/04/index.html}{Big Fish, little pond}\\(2024)} 
& Koi Stealer & 5k & 2.10MB \\

\makecell[l]{\href{https://www.malware-traffic-analysis.net/2024/08/15/index.html}{WarmCookie}\\(2024)} 
& WarmCookie & 18k & 12.00MB \\

\makecell[l]{\href{https://www.malware-traffic-analysis.net/2024/07/30/index.html}{You dirty rat!}\\(2024)} 
& STRRAT (RAT) & 12k & 11.50MB \\

\makecell[l]{\href{https://www.malware-traffic-analysis.net/2022/02/23/index.html}{SunnyStation}\\(2022)} 
& XLoader, Emotet & 30k & 19.80MB \\

\makecell[l]{\href{https://www.malware-traffic-analysis.net/2022/03/21/index3.html}{BurninCandle}\\(2022)} 
& IcedID, Cobalt Strike & 16k & 6.90MB \\

\makecell[l]{\href{https://www.malware-traffic-analysis.net/2022/01/07/index.html}{Spoonwatch}\\(2022)} 
& Oski Stealer & 6k & 4.70MB \\

\makecell[l]{\href{https://www.malware-traffic-analysis.net/2021/09/10/index.html}{Angry Poutine}\\(2021)} 
& BazarLoader & 9k & 6.20MB \\

\makecell[l]{\href{https://www.malware-traffic-analysis.net/2021/02/08/index.html}{Ascolimited}\\(2021)} 
& Hancitor, Ficker Stealer & 21k & 11.10MB \\

\bottomrule
\end{tabular}
\end{table}

\vspace{0.1cm} \textbf{Metrics}: We evaluate the results using quantitative and qualitative metrics, reflecting how a SOC analyst would assess a malware investigation. Quantitatively, we measure the agent’s ability to extract \textbf{victim details} (hostname, IP, and user account) from each report by comparing the agent output against the ground-truth metadata provided in the official exercise solutions. We compute accuracy as the ratio of correctly retrieved fields across all 10 traces. Qualitatively, we perform a manual analysis of the \textbf{Indicators of Compromise (IOCs)} -- including domains, IPs, and hashes -- identified by the agent. We compare the generated IOC list against the reference solution to determine matches, partial matches, and misses. Also, we assess whether the agent uses its integrated web-search component to contextualise IOCs and link them to known malware campaigns or threat intelligence sources. 
\section{Results}\label{sec:results}
In this section, we systematically evaluate design alternatives for building \tool. We use 20 traces from the scenario of attacks against web services to determine the best agent architecture and compare different backend LLMs. Then, we use the 10 events from 2025 vulnerabilities to test the selected agent. Eventually, we conduct a portability study on 10 traces from the scenario of malware traffic analysis. 

For our experiments, we use well-known LLMs exposed through public interfaces.\footnote{We directly use OpenAI API and other LLMs' API using \url{https://www.together.ai/} to simplify the design. The API access method has no impact on the LLM results.} We evaluate four architectures and six different LLMs. We keep the default LLMs' temperatures and repeat three times each incident to account for nondeterminism.\footnote{Regulates the randomness in model generation. Ranges from 0 to 1, where 0 is deterministic generation. Empirical evidence~\cite{lee2022factuality,wang2023selfconsistency} shows that deterministic generations can impair performance.} The time to complete an incident analysis varies from a few to 15-20 minutes based on the incident and LLM. We set the maximum number of steps to 25.

    \subsection{Which Agent Architecture}\label{sec:arch_abl}
    First, we focus on different architectures and fix the back-end LLM to OpenAI GPT-4o.
We summarise results in Tab.~\ref{tab:architecture_ablation}.

\vspace{0.1cm}\textbf{Best Architecture}: To study the effects of improved memory and web-search, we begin comparing \baseline (SA) with the \textit{CFA-bench} baseline. Both approaches achieve similar accuracy in identifying the service under attack. However, thanks to the improved Web Search tool and MemGPT-style memory management, SA achieves substantially better results in other dimensions: CVE identification improves by $+14\%$, and attack success evaluation by $+34\%$. We attribute these gains to SA’s ability to retrieve better targeted information about candidate CVEs from the Web, and to accumulate and reuse contextual information over multiple reasoning steps to refine its conclusions, thanks to its improved memory management.

Despite these improvements, the overall performance of SA remains unsatisfactory. SA needs a large number of steps to solve the problem because it iteratively inspects packets until it finds some evidence. This erratic search causes a waste of input tokens and leads the model to eventually lose focus. Additionally, the MCC is negative: This happens because SA tends to return ``attack unsuccessful'' whenever it fails to gather sufficient evidence. Because the benchmark is unbalanced (14 successes versus 6 failures), this systematically biases SA’s predictions towards the wrong class, explaining its poor MCC despite reasonable accuracy and F1 scores\footnote{Note: a random guess for ‘‘attack success'' would have 50\% accuracy.}.

\begin{table}[h]
\centering
\caption{\textbf{Quantitative metrics of agent architectures on attacks against web services}. FRA delivers the best accuracy and efficiency. Best in \textbf{bold}, second-best \underline{underlined}.}
\label{tab:architecture_ablation}
\footnotesize
\resizebox{1.0\columnwidth}{!}{%
\begin{tabular}{lccccc}
\hline
\multicolumn{2}{c}{\textbf{Metric}} &
  \textbf{\cfabench} & 
  \textbf{SA} &
  \textbf{TEA} &
  \textbf{FRA} \\ \hline
\multicolumn{2}{l}{\textbf{Service ($\uparrow$)}}       & 0.42 & 0.45       & {\ul 0.58} & \textbf{0.67} \\
\multicolumn{2}{l}{\textbf{CVE ($\uparrow$)}}           & 0.14 & 0.28       & {\ul 0.35} & \textbf{0.45} \\
\multicolumn{1}{l|}{\multirow{3}{*}{\textbf{\rotatebox{90}{Success}}}} &
  \textbf{Acc ($\uparrow$)} &
  0.13 &
  0.47 &
  {\ul 0.48} &
  \textbf{0.62} \\
\multicolumn{1}{l|}{}     & \textbf{F1($\uparrow$)}     & ---  & {\ul 0.49} & 0.46       & \textbf{0.62} \\
\multicolumn{1}{l|}{}     & \textbf{MCC ($\uparrow$)}   & ---  & -0.11      & {\ul 0.00}  & \textbf{0.45}  \\
\multicolumn{2}{l}{\textbf{Steps ($\downarrow$)}}       & ---  & 18.11      & {\ul 11.32} & \textbf{5.48}  \\
\multicolumn{2}{l}{\textbf{In. Tokens ($\downarrow$)}}  & ---  & 5.48M      & {\ul 4.37M} & \textbf{3.86M} \\
\multicolumn{2}{l}{\textbf{Out. Tokens ($\downarrow$)}} & ---  & 78k        & 123k        & 112k     \\
\multicolumn{2}{l}{\textbf{Cost [\$] ($\downarrow$)}}   & ---  & 14.48      & {\ul 12.15} & \textbf{10.78}
\end{tabular}%
}
\end{table}

Now, focus on the performance of the \tshark (TEA). The agent outperforms SA, but performance is still limited.
TEA’s main limitation lies in the coordination gap between the main agent and the \textit{Tshark sub-agent}. As shown in Appendix~\ref{sec:tshark_breakdown}, the main agent often issues broad prompts (e.g., “explore the HTTP traffic”), leading the \textit{Tshark sub-agent} to perform inconclusive analyses and miss useful evidence. Still, when the \textit{Tshark sub-agent} extracts a meaningful clue, TEA resolves the task efficiently. It identifies the correct service in 35 of 60 runs and the corresponding CVE in 21 of them. Once the service is known, it finds the CVE within three reasoning steps (query → reasoning → second query). In a nutshell, while TEA enables detailed packet analysis, its overall effectiveness is constrained by weak coordination between its two agent layers.

\begin{figure}[thb]
    \centering
    \includegraphics[width=.8\columnwidth]{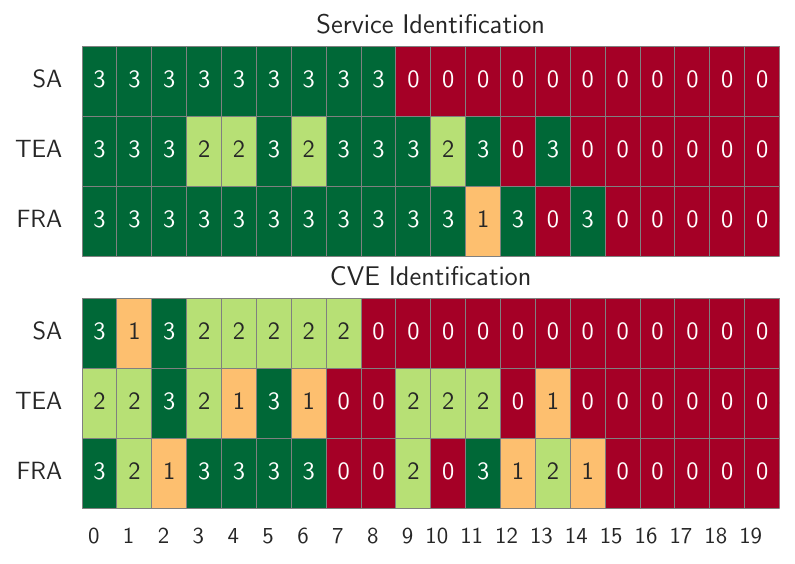}
    \caption{\textbf{Breakdown of correct identifications (top: service, bottom: CVE) over 20 incidents.} FRA outperforms others, both in simpler (left) and harder (right) samples. }
    \label{fig:service_cve_identification}
\end{figure}

Move now to the FRA results. This architecture resolves the coordination gap by decoupling the \textit{Flow Summariser} sub-agent from the main agent. The summariser independently conducts a systematic analysis of the PCAP, producing a high-quality report of suspicious activities without being influenced by intermediate reasoning steps. The main agent then refines these findings through targeted reasoning and web searches. Although this limits the agent’s action space -- it cannot query the summariser again -- FRA achieves the best overall performance at the lowest cost. On average, it converges in just $\sim$5 steps, significantly reducing input token usage (and cost).
Notice finally $MCC>0$, testifying that FRA takes well-grounded decisions based on the evidence collected by the \textit{Flow Summariser} sub-agent.

\vspace{0.1cm}\textbf{Result Breakdown}
Fig.~\ref{fig:service_cve_identification} reports, for each incident in the dataset, how many times (out of three runs) the agents correctly identify the service (top) and the CVE (bottom). Incidents are ordered by difficulty.
SA can correctly identify the service for the first 8 incidents only. In 5 of them, the agent guesses the exact service in the first reasoning step. In fact, for those incidents, the answer is self-contained in the \texttt{packet list summary}. When that is the case, the agent can generally get the correct CVE with 1/2 web searches afterwards. In contrast, when finding the service requires using the \textit{PCAP reader tool}, the reasoning trace becomes much longer (18 steps on average, Table~\ref{tab:architecture_ablation}). The agent starts losing focus and finally fails in both tasks.

TEA shows improved service identification thanks to the \textit{TShark sub-agent}, but the coordination issues discussed earlier prevent it from translating partial findings into successful CVE detection. By contrast, FRA’s decoupling strategy ensures more consistent progress: the Flow Summariser provides focused evidence, enabling the main agent to refine its reasoning and achieve higher accuracy in both service and CVE identification.

Looking at the hardest incidents (15-19), these involve multi-container systems where the attack generates a large volume of traffic (Tab.~\ref{tab:test_set}), causing the agents to lose focus. Incident 17-Cacti is a notable exception: here, the difficulty stems not from traffic overload but from the scarcity of online information. In this scenario, the Web Search tool provides little support, and agents ultimately fail.
Conversely, TEA and FRA fail to identify the CVE in incidents 7 and 8. In these cases, Apache ActiveMQ exposes OS and kernel version details in plain text. The attacker connects via telnet and issues a standard HTTP GET request, producing only faint network traces. In TEA, these subtle signals go unnoticed by the main agent that, as a result, does not trigger the \textit{Tshark sub-agent} to perform a deeper inspection. In FRA, the main agent focuses solely on payload analysis and, therefore, overlooks the telnet traffic entirely (see Section~\ref{sec:limitations}).


\vspace{0.1cm}\textbf{Web Search Tool Usage}: We further analyse how agents leverage the Web Search tool by examining a subset of increasingly complex incidents (Fig.~\ref{fig:websearches}). Each row reports the last query issued in one run, with cell colours reflecting whether the correct CVE was identified (green) or not (red).

We identify three typical failures: (A) query asking for the wrong service, (B) query broadly looking for attacks, and (C) query asking for the wrong attack. At large, a ``correct'' query does not guarantee success, and a wrong query implies failure.
For instance, service misidentification (A) makes the whole analysis fail (see SA looking for \texttt{JetDirect} and \texttt{etcd} in the \textit{Couch DB} and \textit{APISIX} incidents). In the Gitlab incidents, SA gets lost in investigating the long packet list, reaching the maximum step number.
TEA performs slightly better, e.g., correctly looking for \texttt{CouchDB}. Yet, its queries are too broad to succeed in complex scenarios.
Only FRA manages to recover CVEs for some challenging incidents. Curiously, in the \texttt{12-APISIX} incident, queries are correct, and the true CVE is present in the web search results. Yet the agent picks the right CVE in only one run, suggesting that GPT-4o struggles at interpreting web search results.


\begin{figure*}[thb]
    \centering
    \includegraphics[width=0.9\linewidth, trim=20 130 20 130, clip]{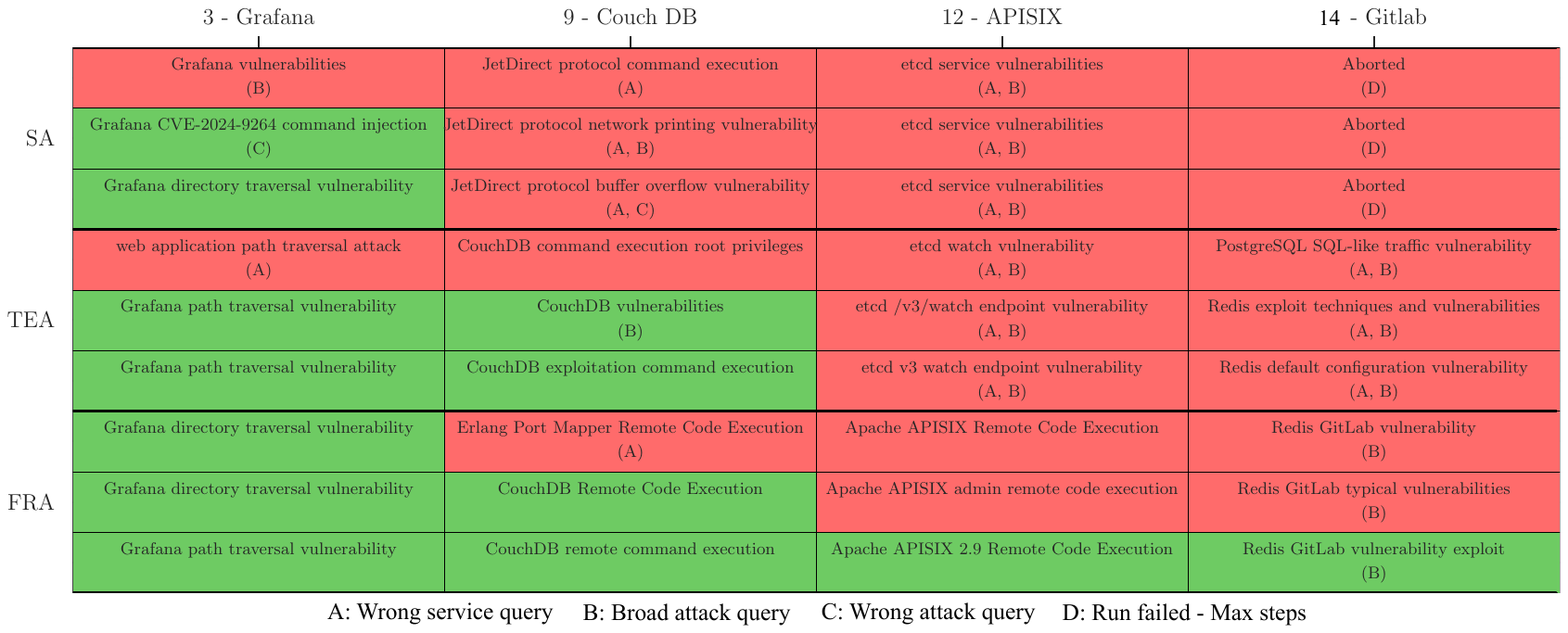}
    \caption{\textbf{Web-search outcomes across incidents, with red cells marking failed CVE identifications.} FRA issues more accurate web queries, succeeding where SA and TEA fail.}
    \label{fig:websearches}
\end{figure*}
    \subsection{Which LLM Backend}\label{sec:llm}
    Now, we study the effects of different back-end LLMs on the performance while providing some quantitative interpretation of their reasoning. We test only FRA -- that proved to be the best architecture -- and compare six LLMs. 

\begin{table}[h]
\centering
\setlength{\tabcolsep}{3.5pt} 
\renewcommand{\arraystretch}{1.1} 
\caption{\textbf{Quantitative Metrics of FRA across six back-end LLMs for attacks against web services.} 
Open-source LLMs rival or surpass proprietary ones while cutting cost. 
Best in \textbf{bold}, second-best \underline{underlined}.}
\label{tab:llm_ablation}
\resizebox{\columnwidth}{!}{%
\begin{tabular}{lccccccc}
\hline
\multicolumn{2}{c}{\textbf{Metric}} & 
\textbf{GPT-4o~\cite{openai2024gpt4technicalreport}} & 
\textbf{o3~\cite{openaio3}} & 
\textbf{GPT-5~\cite{gpt5}} & 
\textbf{\begin{tabular}[c]{@{}c@{}}Deepseek\\ R1~\cite{deepseekr1}\end{tabular}} & 
\textbf{\begin{tabular}[c]{@{}c@{}}Kimi\\ K2~\cite{kimiteam2025kimik2openagentic}\end{tabular}} & 
\textbf{LLaMA-4~\cite{llama4}} \\ \hline

\multicolumn{2}{l}{\textbf{Service ($\uparrow$)}} & 0.67 & 0.75 & 0.80 & \textbf{0.85} & \underline{0.82} & 0.75 \\
\multicolumn{2}{l}{\textbf{CVE ($\uparrow$)}} & 0.45 & 0.48 & \textbf{0.68} & \underline{0.67} & 0.63 & 0.53 \\

\multicolumn{1}{l|}{\multirow{3}{*}{\textbf{\rotatebox{90}{Success}}}} 
  & \textbf{Acc ($\uparrow$)} & 0.62 & \textbf{0.80} & \underline{0.78} & \underline{0.78} & \underline{0.78} & 0.60 \\
\multicolumn{1}{l|}{} & \textbf{F1 ($\uparrow$)} & 0.62 & \textbf{0.79} & \underline{0.77} & 0.76 & 0.76 & 0.63 \\
\multicolumn{1}{l|}{} & \textbf{MCC ($\uparrow$)} & 0.45 & \textbf{0.65} & \underline{0.63} & 0.55 & 0.55 & 0.35 \\

\multicolumn{2}{l}{\textbf{Steps ($\downarrow$)}} & 5.48 & \textbf{3.55} & \underline{3.93} & 5.42 & 6.70 & 17.12 \\
\multicolumn{2}{l}{\textbf{In. Tokens ($\downarrow$)}} & \underline{3.86M} & \textbf{3.67M} & 3.92M & 4.66M & 4.35M & 8.69M \\
\multicolumn{2}{l}{\textbf{Out. Tokens ($\downarrow$)}} & \underline{112k} & 245k & 373k & 555k & \textbf{94k} & 215k \\
\multicolumn{2}{l}{\textbf{Cost {[}\${]} ($\downarrow$)}} & 8.60 & 9.30 & 8.63 & \underline{3.78} & 4.63 & \textbf{2.52} \\ 
\hline
\end{tabular}
}
\end{table}

In Tab.~\ref{tab:llm_ablation} we detail the results. Focusing first on the closed-source OpenAI models, both \texttt{o3} and \texttt{GPT-5} outperform the baseline \texttt{GPT-4o} in service identification (+8\% and +13\%, respectively). This aligns with expectations: \texttt{o3} is a dedicated \textit{reasoning model}, designed to generate an internal reasoning trace before producing the final answer, at the cost of higher latency. This makes it well-suited for detailed forensic investigations. \texttt{GPT-5}, in contrast, is a \textit{hybrid model} that can dynamically invoke reasoning when needed and benefits from a more recent training process.
However, for CVE detection, \texttt{o3} performs only on par with \texttt{GPT-4o}. This happens because \texttt{o3} is often overconfident in its own knowledge and relies less on the Web Search tool (see later). Still, thanks to its reasoning capabilities, \texttt{o3} achieves the highest performance in the evaluation of successful attacks (0.64 MCC), even surpassing the newer \texttt{GPT-5}. Both models also require fewer reasoning steps on average, but their verbosity makes the overall cost comparable to \texttt{GPT-4o}. In Appendix~\ref{sec:web_scraping}, we provide a practical example of how \texttt{GPT-5} can improve the performance of \texttt{GPT-4o}.

Turning to open-source models, both \texttt{DeepSeek R1} and \texttt{Kimi-K2}\footnote{Open-source models make their code, architecture, and trained weights freely available for use, modification, and redistribution.} deliver impressive results, matching or surpassing OpenAI’s best models in most metrics. This is particularly encouraging, as they can be deployed on-premise at inference time, avoiding the need to outsource sensitive data. Even when accessed via the cloud (as in our experiments), these models remain significantly cheaper than their closed-source counterparts -- roughly half the cost.
The open source \texttt{Llama-4 Maverick} improves over \texttt{GPT-4o}, but ranks only second to last overall.

In short, open-source LLMs rival or even exceed the performance of the best proprietary models while offering greater flexibility and substantially lower cost.

\begin{figure}[htp]
    \centering
    \includegraphics[width=.9\columnwidth]{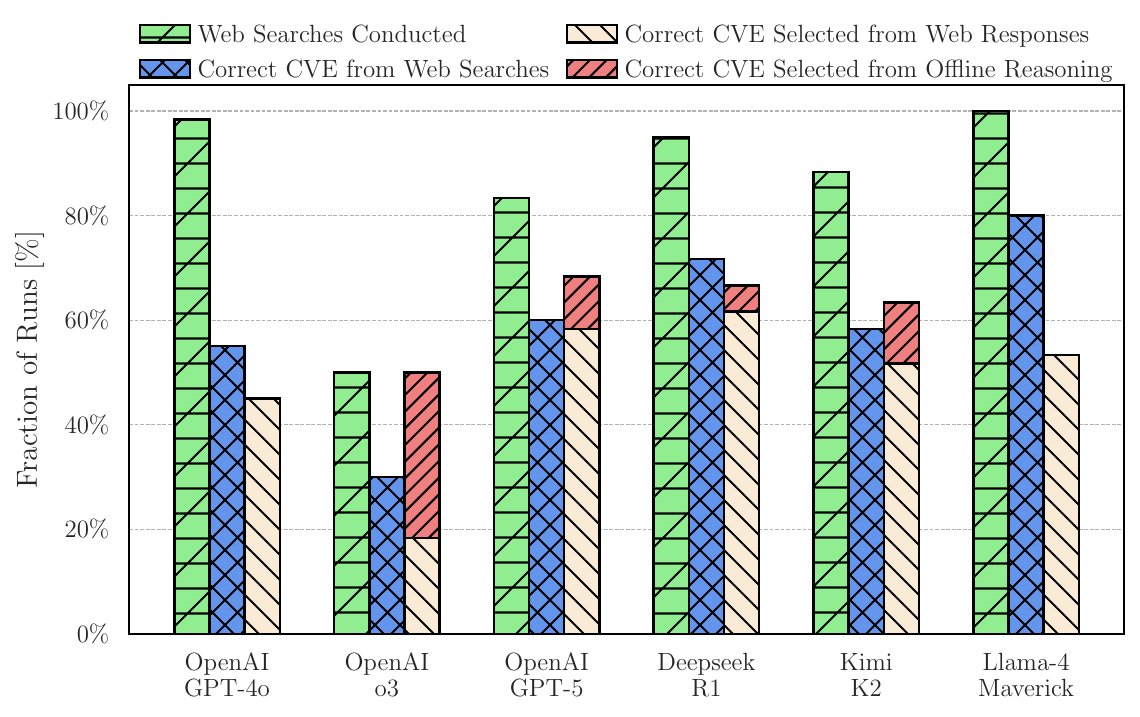}
    \caption{\textbf{Breakdown of web-search behaviour and CVE identification outcomes across 60 runs}. Models differ sharply in how they balance web use and reasoning for CVE detection. The summed detection rates match Table~\ref{tab:llm_ablation}.}
    \label{fig:cve_query}
\end{figure}

\vspace{0.1cm}\textbf{Result Breakdown}: To provide a more interpretable view of the model performance, Fig.~\ref{fig:cve_query} analyses CVE identification outcomes across all 60 runs. For each model, we report the fraction of runs:
i) where the model issues a web query (green bar);
ii) where the last web response contains the correct CVE identifier (blue bar);
iii) the final CVE successfully detected, distinguishing correct answers supported by web evidence (cream bars) from those obtained through prior knowledge (red bar).

All models use the Web Search tool in over 80\% of cases, except \texttt{o3}, which does so in only 50\% of runs. This reflects its distinctive behaviour: despite the prompt explicitly encouraging web searches when uncertain (see Appendix~\ref{sec:prompts}), \texttt{o3} tends to rely heavily on its internal knowledge, often displaying overconfidence.
The quality of the queries is captured by the blue bar. Here, \texttt{GPT-4o} performs poorly: it searches in 95\% of runs, yet only 54\% of responses contain the correct CVE. By contrast, \texttt{Llama-4 Maverick} emerges as the most effective searcher: it always queries the web, and 80\% of its results include the correct CVE.
The crucial step is then selecting the right CVE among candidates. \texttt{GPT-5}, \texttt{DeepSeek R1}, and \texttt{Kimi K2} excel here: whenever the correct CVE is present, they pick it 90-95\% of the time (blue bar $\approx$ cream bar). \texttt{Llama-4 Maverick}, however, struggles to distinguish between near-identical CVEs, choosing the wrong one in 30\% of such cases showing a clear reasoning limitation.
Finally, the red bar highlights cases where the correct CVE is predicted without appearing in web search results, i.e. by relying on LLM's prior knowledge. As said, {o3} frequently succeeds in this way. Interestingly, \texttt{GPT-5} shows a more systematic use of external evidence, only occasionally complementing it with prior knowledge.

Overall, the figure reveals three distinct behaviours: \texttt{o3} relies primarily on offline priors; \texttt{GPT-5} and \texttt{DeepSeek R1} combine web evidence with strong reasoning in a balanced manner; \texttt{Llama-4 Maverick} and \texttt{GPT-4o} suffer from mismatches between retrieval and reasoning.

    \subsection{Test on 2025 CVE and Benign Traces}\label{sec:inference}
    \begin{table}[t]
\centering
\caption{\textbf{Performance of top LLM backends on 2025 incidents.} GPT-5 and o3 achieve the highest accuracy. Best in \textbf{bold}, second-best \underline{underlined}.}
\label{tab:test_set}
\footnotesize
\resizebox{\columnwidth}{!}{%
\begin{tabular}{lcccc}
\multicolumn{2}{c}{\textbf{Metric}}                                                                & \textbf{Deepseek R1} & \textbf{OpenAI o3} & \textbf{OpenAI GPT-5} \\ \hline
\multicolumn{2}{l}{\textbf{Service ($\uparrow$)}}      & \textbf{0.90} & {\ul 0.80}    & \textbf{0.90} \\
\multicolumn{2}{l}{\textbf{CVE ($\uparrow$)}}          & \textbf{0.80} & {\ul 0.70}    & \textbf{0.80} \\
\multicolumn{1}{l|}{\multirow{3}{*}{\textbf{\rotatebox{90}{Success}}}} & \textbf{Acc ($\uparrow$)} & {\ul 0.70}           & \textbf{0.90}      & \textbf{0.90}        \\
\multicolumn{1}{l|}{}    & \textbf{F1($\uparrow$)}     & {\ul 0.73}          & \textbf{0.90} & \textbf{0.90}    \\
\multicolumn{1}{l|}{}    & \textbf{MCC ($\uparrow$)}   & {\ul 0.41}    & \textbf{0.82} & \textbf{0.82} \\
\multicolumn{2}{l}{\textbf{Steps ($\downarrow$)}}      & 5.5           & \textbf{4.3}  & {\ul 4.4}     \\
\multicolumn{2}{l}{\textbf{In. Token ($\downarrow$)}}  & {\ul 492k}    & \textbf{298k} & 502k          \\
\multicolumn{2}{l}{\textbf{Out. Token ($\downarrow$)}} & 207k          & \textbf{108k} & {\ul 185k}    \\
\multicolumn{2}{l}{\textbf{Cost [\$] ($\downarrow$)}}  & \textbf{0.72} & {\ul 1.46}    & 2.47         
\end{tabular}%
}
\end{table}

\vspace{0.1cm}\textbf{CVEs from 2025}: We finally evaluate \tool on 10 recent CVEs and benign traffic, using {DeepSeek R1}, {OpenAI o3}, and {GPT-5} as back-end LLMs. As shown in Tab.~\ref{tab:test_set}, the agents achieve 90\% service identification and 80\% CVE detection accuracy -- an impressive result considering that these incidents involve both vulnerable and patched systems. {GPT-5} and {o3} deliver the strongest performance, while {DeepSeek R1} struggles with attack success classification (3 failures out of 10), yielding an MCC of only 0.41, suggesting a more random guess approach.
These results support the choice of GTP-5 as the final backend LLM for \tool.

\vspace{0.1cm}\textbf{Benign traces}: Finally, challenging \tool with 10 benign traces collected during simple browsing sessions, it correctly reports no evidence of malicious activity. In two cases, it flags potential brute-force attempts, supported by repeated and rapid failed login events, which, in fact, is an accurate observation.

    \subsection{Human Evaluation}\label{sec:human_eval}
    In this section, we conclude our analysis on the attacks on web-services by showing the result on the evaluation of the reports generated by OpenAI o3 and DeepSeek R1 on four successful attacks in the 2025 testset.
25 volunteers, including students, experienced researchers and professors, responded. 12 participants self-declared low (rating 1-2), 8 medium (rating 3), and 5 high expertise (rating 4-5).

\begin{figure}[t]
    \centering
    \includegraphics[width=0.8\columnwidth]{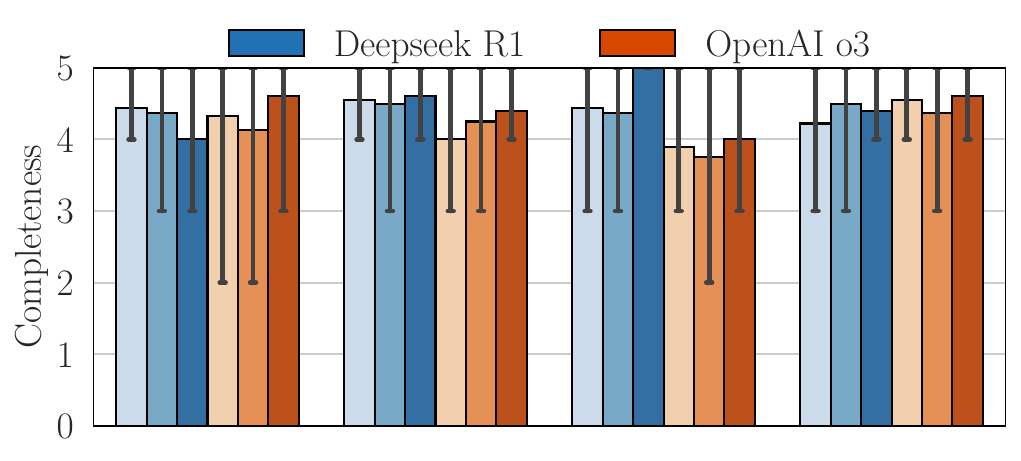}\\
    \includegraphics[width=0.8\columnwidth]{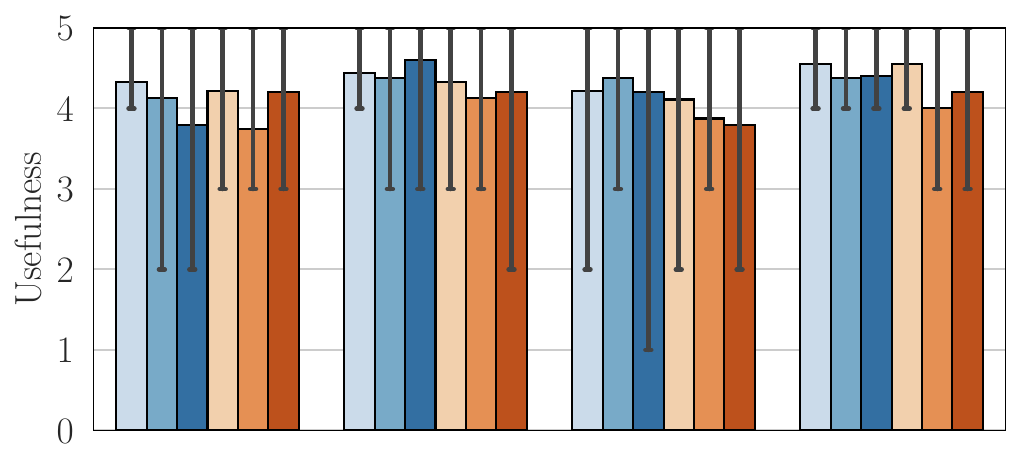}\\
    \includegraphics[width=0.8\columnwidth]{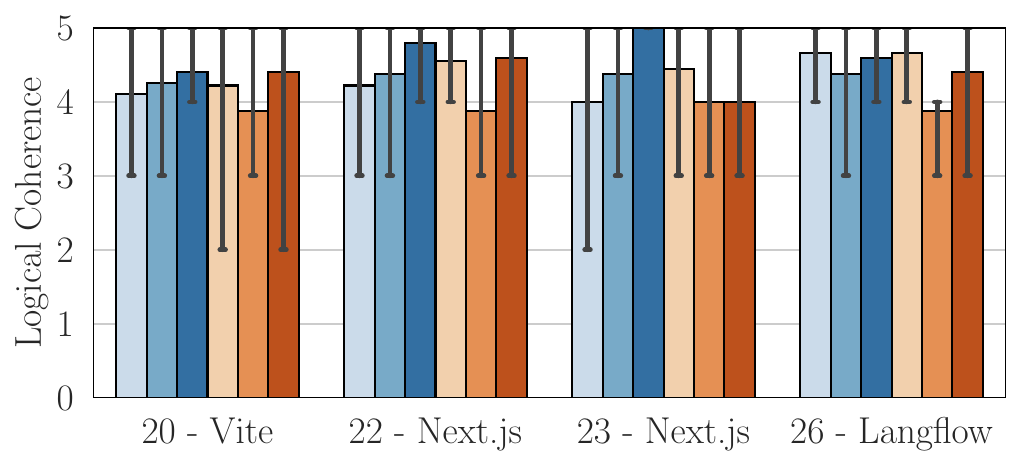}
    \caption{\textbf{Average scores for Completeness, Usefulness, and Logical Coherence by incident and expertise level.} Both o3- and DeepSeek-based agents are rated highly across all evaluation criteria. Whiskers show min–max range.}
    \label{fig:human_scores}
\end{figure}

All participants find the produced reports to be complete, useful, and logically coherent (average score of 4.33, 4.23, 4.31, respectively). In Figure~\ref{fig:human_scores}, we detail the average, minimum, and maximum scores for each evaluation criterion. There are only minor variations between the different levels of expertise and between the two LLMs.

When asked which report they preferred, participants showed a slight overall preference for Deepseek R1, particularly among medium and high experience. This is reflected by the all-metric average grade for Deepseek R1 of 4.39 versus 4.20 for o3.
Participants preferred OpenAI o3 only on the first incident (20 – Vite), as they reported that Deepseek R1 comment was too verbose in this case. In the other cases, they also found OpenAI o3 to be overconfident.

    \subsection{Portability Study: Malware Traffic Analysis}\label{sec:malware_analysis}
    
Next, we assess the portability of \tool by applying it to malware traffic analysis. The goal is to determine whether design choices that prove effective for analysing attacks on web services generalise to a distinct forensic scenario -- still centred on PCAP analysis.

For this, we modify \tool by changing only the task description prompt:
\begin{itemize}
    \item \textbf{Main Prompt} -- we revise it to request an executive summary including victim details (hostname, IP address, Windows username) and Indicators of Compromise (IOCs).
    \item \textbf{Web Search Tool} -- we adapt it to enable (i) the retrieval of public information on vulnerabilities or exploit techniques, (ii) validation of known IOCs, and (iii) the identification of associated malware families or campaigns. Web queries serve solely to enrich, not replace, observed evidence.
\end{itemize}
All other components remain unchanged. 

We instruct the modified \tool to analyse the 10 malware traces and manually check the results. The results are encouraging: In 10 malware infections, the agent correctly identifies victim details in 9 cases and consistently extracts the most relevant IOCs (see Appendix~\ref{app:detail_malware_analysis}). It also recognises activity patterns linked to known criminal groups (e.g., CobaltStrike, NetSupport Rat), confirming them via the web search tool.

Although the agent does not always report exhaustive command-and-control details -- by design, these were out of scope -- the overall performance indicates strong portability. The core design principles behind \tool generalise effectively to other forensic contexts, requiring only minimal adaptation through prompt engineering.
\section{Discussion and Limitations}\label{sec:limitations}
We discuss the current limitations of our architecture and provide a possible future exploration of this topic.


\vspace{0.1cm}\textbf{Benchmark Scope and Realism}: Although the benchmark consists of traces from two scenarios describing heterogeneous network incidents and malware analysis, these traces are produced within containerised environments or controlled test beds, rather than captured from live enterprise networks. As a result, they may lack the unpredictability, background noise, and concurrent benign activity typical of operational settings. Most incidents target known vulnerabilities or observable behaviours. While this design supports systematic and reproducible evaluation, it limits the generalisability of the findings, simplifying the complexity of real-world forensic scenarios. Future work should encompass larger and more diverse datasets -- including anonymised real-world traffic -- to better assess scalability and robustness in real-world scenarios.
Similarly, we will investigate what other input data could be considered to improve the processing of evidence.


\vspace{0.1cm}\textbf{Balancing Flexibility and Reliability}: Network incidents require forensic agents to flexibly search for evidence in heterogeneous data. To approximate human practice, the initial system, \tshark (TEA), allowed flexible use of `tshark' commands to analyse traffic and gather evidence, mimicking human investigation. However, this flexibility often caused the agent to lose focus or misuse commands. To address these problems, we developed the \constrain (FRA) agent, which limits the analysis of connection payloads using a summarisation sub-agent. Although this improved reliability, it also reduced the agent's capability to detect evidence in transport-layer behaviours (e.g., TCP resets in SSH attacks). To address this limitation, future work should combine FRA’s efficiency with TEA’s flexibility, e.g., by training a dedicated \texttt{tshark}-expert sub-agent that can execute and validate commands safely.


\vspace{0.1cm}\textbf{Feedback Loop for Self-Correction}: In our best-performing architecture (FRA), the main agent investigates incidents sequentially, starting from the pre-processed summary, without the ability to revisit the raw data. This contrasts with real-world practice, where experts collaborate and re-examine raw evidence, thus limiting the agent's capability. This limitation reflects the complexity and challenges of tightly orchestrating multiple agents~\cite{laban2025lost} -- especially over long interactions with evolving hypotheses -- which reduces opportunities for self-correction. Future work should explore orchestration strategies that enable reflective feedback loops (e.g., cross-agent critique, mediated tool calls) without sacrificing reliability, whether through more advanced architectures or improved prompting. In addition, iterative analysis challenges the LLM's natural ability to process large amounts of data. Therefore, splitting the data and analysis into chunks may impair the agent's ability to reach correct conclusions~\cite{son-etal-2024-multi-task} -- as also noted in~\cite{cfabench}.

\section{Conclusion}
In this paper, we introduced \tool, an autonomous LLM-based agent for post-mortem cyber forensics. By systematically exploring agent architectures, memory mechanisms, and language models, we demonstrated that agentic AI can effectively support analytical reasoning and evidence correlation in cybersecurity investigations.
Our findings highlight three key design principles: (i) multi-agent specialization enables sustained reasoning and improves task focus; (ii) simple orchestration outperforms deeply nested coordination schemes; and (iii) design choices generalize across forensic domains, transferring effectively from web-service exploitation to malware traffic analysis with minimal adaptation.
CyberSleuth achieves up to 80\% accuracy on previously unseen 2025 incidents, producing forensic reports rated by experts as coherent, complete, and operationally useful. These results substantiate the feasibility of autonomous, reasoning-capable agents in complex, real-world security workflows.

Beyond its forensic scope, this study offers general insights into the design and evaluation of agentic systems for cybersecurity automation, making it a valuable reference for both practitioners and researchers exploring AI-driven reasoning in security operations.
%

\section*{Appendix}
\label{sec:appendix}
\section{Breakdown of Tshark Expert Agent}\label{sec:tshark_breakdown}

In the following, we present two reasoning tracks generated by the TEA architecture for the event \textit{04 – Apache HTTP Server}. This case is illustrative, as it highlights i) how the main agent can reliably reach the correct solution when the Tshark Expert accurately interprets its instructions and ii) how early miscoordination can hinder further progress. Steps executed by the main agent are marked with \textbf{[MA]}, while those of the Tshark Expert are marked with \textbf{[TE]}. 

\vspace{0.1cm}\textbf{Good coordination:}

\begin{enumerate}[leftmargin=*, itemsep=2pt, topsep=2pt]
\item \textbf{[MA]}: The PCAP summary shows multiple TCP conversations on port~80, indicating HTTP traffic. Asks the Tshark Expert to extract HTTP requests and headers.
\item \textbf{[TE]}: Inspects HTTP requests and observes potential directory traversal attempts. The responses include a mix of \texttt{200} and \texttt{403}. The server banner reveals \texttt{Apache/2.4.49}.
\item \textbf{[MA]}: With the service and attack type identified, it issues a web query: \emph{``Apache 2.4.49 directory traversal command execution vulnerability''}.
\item \textbf{[MA]}: The search returns two candidate CVEs: \texttt{CVE-2021-41773} (Apache 2.4.49) and \texttt{CVE-2021-42013} (Apache 2.4.50). Despite already having the correct version, \textbf{MA} requests deeper evidence: it asks \textbf{TE} to highlight HTTP requests showing directory traversal or unusual file access patterns.
\item \textbf{[TE]}: Attempts three times to filter requests containing \texttt{../} via simple string matching, but the filters fail each time.
\item \textbf{[MA]}: Instructs \textbf{TE} to refine the regular expression and retry.
\item \textbf{[TE]}: The refined filter succeeds; however, focusing on requests alone does not further disambiguate the CVE.
\item \textbf{[MA]}: Concludes by linking the observed banner \texttt{Apache/2.4.49} to \texttt{CVE-2021-41773}. Given the presence of \texttt{200} responses to traversal attempts, it labels the attack as \emph{successful}.
\end{enumerate}

\vspace{0.1cm}\textbf{Bad coordination:}

\begin{enumerate}[leftmargin=*, itemsep=2pt, topsep=2pt]
  \item \textbf{[MA]}: Observes multiple TCP conversations on port~80, inferring HTTP traffic. Requests the extraction of HTTP requests and headers to identify the service and possible attack patterns.
  \item \textbf{[TE]}: Extracts HTTP requests but finds limited metadata (empty \texttt{User-Agent}, no \texttt{Referer} or \texttt{Cookie}). Does detect suspicious directory traversal attempts (\texttt{../}) but fails to clearly identify the service. 
  \item \textbf{[MA]}: Notes evidence of directory traversal and requests further analysis of HTTP URIs to better identify the targeted service.
  \item \textbf{[TE]}: Extracts several suspicious URIs accessing sensitive files (e.g., \texttt{/etc/passwd}, \texttt{/etc/shadow}, \texttt{/var/log/syslog}). Attempts to correlate streams with conversations, but errors arise due to incorrect command syntax. Despite this, it confirms strong signs of directory traversal.
  \item \textbf{[MA]}: Proceeds with a web search using the generic query \emph{``web service directory traversal vulnerability''}.
  \item \textbf{[MA]}: Receives multiple candidate CVEs across unrelated products (Cisco ASA, Barracuda appliances, WordPress plugins, etc.). Picks \texttt{CVE-2020-3452} (Cisco ASA) as the most plausible match.
  \item \textbf{[MA]}: Requests further evidence by asking the Tshark Expert to extract HTTP responses for the suspicious requests.
  \item \textbf{[TE]}: Extracts responses with status code \texttt{200}, revealing leaked sensitive data from \texttt{/etc/passwd}, \texttt{/etc/hosts}, and other files. Confirms that the directory traversal attack succeeded.
  \item \textbf{[MA]}: Concludes the analysis: identifies \texttt{CVE-2020-3452} as the most relevant CVE, labels the affected service as Cisco ASA, and marks the attack as successful.
\end{enumerate} 
\section{Better LLMs Improve Web-Scraping Accuracy}\label{sec:web_scraping}

We provide an illustrative example showing how stronger reasoning abilities of newer LLMs improve web-scraping relevance and, consequently, incident analysis accuracy. In particular, we focus on the event \textit{12 -- Apache APISIX}. We use a Flow Reporter Agent (FRA) with two different backend models: GPT-4o and GPT-5. Both models retrieve the same two candidate CVEs for the incident: CVE-2021-45232\footnote{\url{https://nvd.nist.gov/vuln/detail/CVE-2021-45232}} and CVE-2022-24112\footnote{\url{https://nvd.nist.gov/vuln/detail/CVE-2022-24112}}. 

\noindent\textbf{GPT-4o}: the information recovered is too broad and not specifically tailored to the observed traffic. 

\begin{itemize}[leftmargin=*, itemsep=2pt, topsep=2pt]
\item \textbf{CVE-2021-45232}: This vulnerability affects the Apache APISIX Dashboard, enabling unauthorized access that leads to remote code execution (RCE). It is critical for analyzing systems using Apache APISIX since unauthorized users could potentially execute arbitrary code, compromising the service or application security. This highlights the significance of scrutinizing any system using APISIX for potential breach indicators.
\item \textbf{CVE-2022-24112}: This CVE pertains to a Remote Code Execution vulnerability in Apache APISIX versions prior to 2.10.0. It is highly relevant for forensic analysts investigating unauthorized execution incidents involving APISIX 2.9.0, as the exploit could allow attackers to execute arbitrary code on the server. This vulnerability underscores the importance of assessing potential breaches or unauthorized access within network infrastructure using Apache APISIX for managing APIs.
\end{itemize}

As a result, the model cannot disambiguate between the two and produces the wrong final answer in two runs.

\noindent\textbf{GPT-5}: The agent successfully links CVE-2021-45232 with forensic evidence -- particularly to the \texttt{/apisix/admin/migrate/import} endpoint. 

\begin{itemize}[leftmargin=*, itemsep=2pt, topsep=2pt]
\item \textbf{CVE-2021-45232}: Apache APISIX Dashboard (Manager API) exposed the \texttt{/apisix/admin/migrate/import} endpoint without proper authentication in vulnerable versions (up to 2.10.1). An unauthenticated attacker could POST a crafted configuration to import routes and enable Lua/script/serverless plugins that APISIX executes, resulting in unauthenticated remote code execution on APISIX nodes. This directly matches the “Manager API migrate import” RCE vector under analysis.
\item \textbf{CVE-2022-24112}: Apache APISIX \texttt{batch-requests} plugin could be abused for server-side request forgery to the Admin API, bypassing IP-based restrictions and enabling unauthorized Admin API operations. Attackers could leverage this to create routes or enable dangerous plugins (e.g., Lua/script) and achieve remote code execution. Often seen in combination with CVE-2021-45232 (e.g., enabling or configuring the plugin via migrate/import) but not applicable in this case, since no evidence of the \texttt{batch-requests} plugin was observed in the traffic.
\end{itemize}

This allows the model to correctly identify the vulnerability in all three runs. 
\section{Prompts}\label{sec:prompts}

We report the \textbf{prompts} we used for our agents.

\vspace{0.1cm}$\bullet$ \textbf{Single Agent}:

\begin{prompt} % DO NOT CHANGE SPACING
Role: You are a specialized network forensics analyst.
You are working towards the final task on a step by step 
manner.

Instruction:
You are provided with a queue of the most recent steps in the 
reasoning process. Store the most relevant information as soon 
as you get them, because the queue has a limited size and 
older messages will be removed.
By thinking in a step by step manner, provide only one single 
reasoning step in response to the last observation and the 
action for the next step. You have tools available to search 
online for CVEs and extract frames from the pcap file.
When you are ready to provide the final answer, stop the 
reasoning and format the result.

Context: Analyze the provided PCAP (Packet Capture) file to 
accomplish the following tasks:

1. Identify the name of the service or application involved.
2. Determine the relevant CVE (Common Vulnerabilities and
Exposures) based on the captured data.
3. Gather evidence of malicious activities associated with
the identified CVE.
4. Assess whether the service or application is vulnerable to
the identified attack.
5. Evaluate whether the attack was successful.

Guidelines:
- Begin by thoroughly analyzing the PCAP file to extract 
relevant information. 
- After the exploratory analysis, ensure that the CVE 
identification is accurate by cross-referencing details from 
external sources with the evidence found in the PCAP files.
- Use the online search tool only after the exploratory 
analysis has been completed to verify the findings and gather
additional information.
IMPORTANT: avoid repeating the same query to the web search 
tool multiple times, the result won't change.

Pcap content: {pcap_content}

{memories}

Queue of steps: {queue}

\end{prompt}

\vspace{0.1cm}$\bullet$ \textbf{Tshark Expert Agent}:

\begin{prompt} % DO NOT CHANGE SPACING
Role: You are a specialized network forensics analyst 
reasoning step by step to investigate an attack against a 
web service, based on a filtered PCAP file and log files. 
Your goal is to determine what happened by correlating expert
analyses, reviewing past reasoning steps and using external
verification tools.

Scenario:
An attacker attempted to exploit a vulnerability in a specific 
web service. All network traffic is filtered to focus on the 
relevant service. The service may or may not be vulnerable, 
and the attack may or may not have succeeded.

Instruction:
You are provided with a queue of the most recent steps in the
reasoning process and a summary of the pcap file given by 
the command: tshark -r <pcap_file> -q -z conv,tcp. 
This command shows you tcp flows to give you an overview of 
the traffic that has to be analyzed more in depth.
The PCAP file is filtered on the traffic of the service of 
interest.
You have access to a specialized sub-agent, the 
"tshark_expert", whose role is to assist you in analyzing 
network traffic through the execution of tshark commands.

The tshark_expert can be called to execute an high 
level analysis on the pcap file, but it has no knowlege 
of CVEs, vulnerabilities or exploits. Specify in the task 
every step in human language, it will translate it into tshark 
commands.
Examples:
1- "Follow the TCP flow number x in the pcap file"
2- "Extract only HTTP headers from tcp stream 2"
etc.
DO NOT CALL THE WEB SEARCH TOOL AND THE TSHARK COMMAND TOOL IN 
THE SAME STEP. WEB SEARCH TOOL CALLS WILL BE IGNORED IN THAT
CASE. 
Store the most relevant information as soon as you get them, 
because the queue has a limited size and older messages will 
be removed.
By thinking in a step by step manner, provide only one single 
reasoning step in response to the last observation and the 
action for the next step. When you are ready to provide the 
final answer, stop the reasoning and format the result.

Context: Analyze the provided PCAP (Packet Capture) file to 
accomplish the following tasks:

1. Identify the name of the service or application involved.
2. Determine the relevant CVE (Common Vulnerabilities and 
Exposures) based on the captured data.
3. Gather evidence of malicious activities associated with 
the identified CVE.
4. Assess whether the service or application is vulnerable
to the identified attack.
5. Evaluate whether the attack was successful.

Guidelines:
- Begin by thoroughly analyzing the PCAP file to extract 
relevant information. 
- Once you understood the service involved and the type of
attack attempted, search for the CVE on the web.
- After the exploratory analysis, ensure that the CVE 
identification is accurate by cross-referencing details 
from external sources with the evidence found in the 
PCAP files.
- Use the online search tool only after the exploratory 
analysis has been completed to verify the findings and 
gather additional information.

Pcap summary: {pcap_content}

{memories}

Queue of steps: {queue}

\end{prompt}

\vspace{0.1cm}$\bullet$ \textbf{Tshark Expert}:

\begin{prompt} % DO NOT CHANGE SPACING
Role: You are an expert in executing tshark commands.
You are working towards completing the assigned task in a 
step-by-step manner.
Use your prior knowledge and tshark manuals if your knowledge
is not enough.
You are capable of translating raw payloads to ASCII and vice 
versa on your own. 
Do not rely on tshark or manual searches for this task -
perform the conversion yourself when needed.

When referring to the tshark manual: only use it to look up 
valid command, filter options or correct syntax. No other 
information can be found on the manual. 

Instructions:
I will provide you with a high-level analysis goal required by 
a forensic analyst on a PCAP file that is already pre-filtered 
on the traffic of a specific service. You are also provided 
with a summary of the PCAP file given by the command: 
tshark -r <pcap_file> -q -z conv,tcp 
that shows you all tcp flows and their order.

General guidelines:
- Every action you take must be guided by the forensic task. 
Do not explore the PCAP randomly without purpose.
- If the output of a command is too large, modify the command 
to filter or adjust the output appropriately before 
executing it.
- If the syntax of the command is correct but there is 
no output for that command, report it as the final result.
- Always prioritize commands that bring you closer to 
answering the forensic task.

You are provided with the previous steps performed during the 
analysis.
By thinking step-by-step, provide only one single reasoning 
step in response to the last observation, followed by the 
action for the next step.

Handling errors:
- If no output is found for a command (even after 
corrections), it is acceptable to report this as the final 
result.
- If the final command you are returning generates an error, 
do not return the specific error but just 'Error in the 
command'.

IMPORTANT:
APPLY A FILTER ONLY AFTER YOU HAVE TRIED TO EXECUTE THE 
COMMAND WITHOUT IT AND THE OUTPUT WAS TOO LONG. 
Moreover, the -c option 
trims the number of packets on which the search is performed, 
it should not be used.
When ready to provide the final answer, call the final answer 
formatter tool.

PCAP summary: {pcap_content}
Task:
{task}
Steps in the analysis:
{steps}

\end{prompt}

\vspace{0.1cm}$\bullet$ \textbf{Flow Reporter Agent}:

\begin{prompt} % DO NOT CHANGE SPACING
Role: You are a specialized network forensics analyst 
reasoning step by step to investigate an attack against 
a web service, based on a filtered PCAP file and log files. 
Your goal is to determine what happened by correlating expert
analyses, reviewing past reasoning steps and using external 
verification tools.

Scenario:
An attacker attempted to exploit a vulnerability in a 
specific web service. All network traffic is filtered to 
focus on the relevant service. The service may or may 
not be vulnerable, and the attack may or may not have 
succeeded.

You are provided with:
1- A FIFO queue containing the most recent reasoning steps. 
Store relevant information early in your reasoning, because 
the queue has a limited size.
2- A forensic report from a PCAP flow analyzer, which analyzed 
each tcp flow in the PCAP file indipendently.
3- Tools to search online and to store relevant information 
in a memory database.

To perform effective web searches, always specify the affected 
service and version (if known) and the identified attack type.

By thinking in a step by step manner, provide only one single 
reasoning step in response to the last observation and the 
action for the next step.When you are ready to provide the 
final answer, stop the reasoning and format the result.

Your task is to analyze the following information and answer 
the four key questions below:

1. What is the name of the service or application involved?
2. What CVE (Common Vulnerabilities and Exposures) is 
associated with the attack, based on expert findings and 
web searches?
3. Was the service vulnerable to this attack?
4. Did the attack succeed?

Pcap flows analysis: {pcap_content}

{memories}

Queue of steps: {queue}

IMPORTANT:
- Don't search on the web for the same or similar queries many 
times. Repeat the web search only if it refers to another 
service or a different attack.

\end{prompt}

\vspace{0.1cm}$\bullet$ \textbf{Flow Summariser}:

\begin{prompt} % DO NOT CHANGE SPACING
You are an expert in analyzing the TCP flow of a PCAP file. 
You analyze one TCP flow at a time to detect suspicious or 
malicious activities against specific services.

You assist a forensic analyst investigating an incident 
involving possible exploitation of a vulnerable service. 
Traffic is filtered for that service, and may include 
related services.

You will receive:
- The report of the analysis done on the previous TCP flows so
that you can correlate findings based also on what happened 
before;
- A chunk containing the text of the TCP flow to be analyzed.

You must:
1. Determine if the traffic indicates an attempted or
successful attack.
2. Identify the targeted service and version, if possible.
3. Specify the type of exploitation 
(e.g., RCE, privilege escalation etc.).
4. Include all relevant observations, such as service responses, 
to help the analyst correlate evidence.

You must produce your output in the following textual format 
(use these field names literally, in English):

Service: [describe the service(s) involved and, if possible, 
their version. Use a comma-separated list if multiple services]

Relevant Events: [describe what the IP addresses involved 
are doing in this TCP flow. Report relevant activities and 
their meaning]

Malicious Activities: [if any suspicious or malicious activity
is found, describe it here and indicate the service affected. 
Otherwise, write "None"]

Attack Success: [indicate whether the attack in this flow 
(or a previous one) appears to be successful or not. If unknown
or not applicable, write "None"]

Be concise and strictly technical. If nothing relevant 
is found, clearly state so in the appropriate fields. Do 
not include extra commentary or formatting.

Analysis of the previous TCP flows with related reports:
{previous_tcp_traffic}

Analysis of the current flow:
Current flow: 

{current_stream}

Chunk:

{chunk}
\end{prompt}

\vspace{0.1cm}$\bullet$ \textbf{Log Summariser}:

\begin{prompt}
You are a forensic AI assistant specialized in log analysis.

Your goal is to assist a forensic analyst by reviewing logs 
related to a specific service under investigation. 
Your analysis will be used in conjunction with full packet 
captures (PCAPs) of the service traffic.

Your task is to:
1. Summarize the most relevant and suspicious events.
2. Identify the service involved and, if possible, its version.
3. Highlight entries suggesting attacks or abnormal behaviors.
4. Include excerpts from the logs to support your findings.
5. Clearly state if no relevant findings are detected.

Tone: concise, technical, forensic.
Audience: a forensic analyst.
IMPORTANT: do not make assumptions on the possible CVE code.

Here is the log content:

{log_content}

Write your report in this format:
Report of the log analysis done by the log reporter: 
<your_report>

\end{prompt}

\section{Malware Traffic - Full IOC results}\label{app:detail_malware_analysis}

Here, we provide a detailed overview of \tool' results when retrieving Indicators of Compromise (IOC) in the analysis of malware traffic.

\begin{table*}[tb]
\centering
\caption{\textbf{Results for the Indicator of Compromise (IOC) detection in the malware traffic scenario}. The model correctly retrieves all the main Indicators of Compromise with respect to the ground-truth and successfully links evidence to \textit{well-known malware attempts}. We manually evaluate responses and compare them with the available ground truth. }
\label{tab:benchmarks}
\footnotesize
\begin{tabular}{p{0.18\textwidth}p{0.41\textwidth}p{0.31\textwidth}}
\toprule
\textbf{Event Name (Year)} & \textbf{Identified IOCs / Evidence} & \textbf{Missed IOCs / Issues} \\ 
\midrule
\href{https://www.malware-traffic-analysis.net/2025/01/22/index.html}{Fake Download Site} (2025) & Recognizes the C2 attempt and the main C2 IP address, and references a fake authenticator site. & Does not report the remaining C2 IPs. \\ 
\midrule
\href{https://www.malware-traffic-analysis.net/2024/11/26/index.html}{Nemotodes} (2024) & Detects \textit{NetSupport RAT} activity, describes the HTTP POST beaconing behavior, and recognizes C2 activity consistent with remote access. & Hallucinates the specific C2 IP, and does not mention the related domains or compromised sites. \\ 
\midrule
\href{https://www.malware-traffic-analysis.net/2024/09/04/index.html}{Big Fish, little pond} (2024) & Detects the \textit{FedEx-themed} download and its source IP, identifies HTTP GET/POST C2 attempts, and reports unusual cookies and an outdated browser version. Notes a large binary HTTP POST linked to possible exfiltration. & Does not report finer-grained details (e.g., full GET/POST attempts and exact URLs). \\ 
\midrule
\href{https://www.malware-traffic-analysis.net/2024/08/15/index.html}{WarmCookie} (2024) & Correctly identifies \textit{STRRAT} as the malware family, reports detailed C2 behavior (e.g., \texttt{ping|STRRAT} marker), detects use of \texttt{ip-api.com/json} for external IP lookup, and mentions GitHub for download/staging. & Hallucinates the C2 IP. \\ 
\midrule
\href{https://www.malware-traffic-analysis.net/2024/07/30/index.html}{You dirty rat!} (2024) & Identifies \textit{KOI Stealer} as the malware family, and mentions that infection is only suspected with no clear signs. & Does not include specific C2 IPs or URLs, and misses the possibly compromised destination URL. \\ 
\midrule
\href{https://www.malware-traffic-analysis.net/2022/02/23/index.html}{SunnyStation} (2022) & Detects the download payload carrying the infection, recognizes subsequent beaconing and C2 traffic, and links them to \textit{IcedID} C2 entries. Identifies suspicious connections consistent with \textit{Cobalt Strike} staging, and describes persistence behaviors aligned with \textit{RAT/Cobalt Strike} workflows. & Does not report finer-grained details (e.g., domain names tied to the IcedID infrastructure or file-sharing domains such as \texttt{filebin.net}, \texttt{situla.bitbit.net}). \\ 
\midrule
\href{https://www.malware-traffic-analysis.net/2022/03/21/index3.html}{BurninCandle} (2022) & Correctly identifies the C2 IP, detects HTTP POSTs masquerading as JPEGs that return Windows PE executables (MZ/PE), and recognizes data exfiltration patterns (large outbound traffic, $\sim$3.1 MB) linked to the \textit{Oski Stealer} profile. Mentions malware artifacts (browser data theft, ZIP structures, cookies, passwords, autofill). & Does not report the full C2 endpoints or reference SHA256 hashes tied to the payloads. \\ 
\midrule
\href{https://www.malware-traffic-analysis.net/2022/01/07/index.html}{Spoonwatch} (2022) & Correctly identifies the malicious domain for payload retrieval, references \textit{BazarLoader/BazarBackdoor} delivery infrastructure, and notes C2 behavior (suspicious outbound TLS sessions). Identifies \textit{Kerberos} ticket abuse (Golden Ticket forgery) and lateral movement attempts. Correctly links the activity to the \textit{TA551 (Shathak)} campaign. & Does not specify the exact IP for the \textit{BazarLoader} download or the detailed \texttt{/bmdff/...} GET path, and omits the specific C2 IPs/domains. \\ 
\midrule
\href{https://www.malware-traffic-analysis.net/2021/09/10/index.html}{Angry Poutine} (2021) & Detects a malicious download using PDF masquerading techniques with embedded MZ/PE content, and links the C2 POST to known attempts (\textit{UA Ghost}, \textit{Header Cowboy}). Notes public IP lookup via \textit{api.ipify.org}. & Does not report the full list of C2 IPs or the specific POST URI path. \\ 
\midrule
\href{https://www.malware-traffic-analysis.net/2021/02/08/index.html}{Ascolimited} (2021) & Links the retrieved IP and domain to \textit{Hancitor}-style C2 behavior, reports persistent C2 activity on port 8080 with \texttt{/ca} and \texttt{/submit.php} endpoints, and identifies payloads associated with \textit{BazarLoader/Cobalt Strike}. Notes exfiltration-related IPs linked to \textit{Ficker Stealer}, public IP lookup via \textit{api.ipify.org}, and correlates the activity chain with typical \textit{Hancitor}/stealer tradecraft. & Does not report the full list of C2 and staging IPs or domains. \\ 
\bottomrule
\end{tabular}
\end{table*} 


\bibliographystyle{plain}
\bibliography{main.bib}

\end{document}